\documentclass[apsprl,twocolumn,superscriptaddress]{revtex4-2}

\usepackage{graphicx,subfigure}
\usepackage{changes}
\usepackage{amsmath,amssymb}
\usepackage{mathtools}
\usepackage{multirow}
\usepackage{textcomp}
\usepackage{float}
\usepackage{xcolor}
\usepackage{titletoc}
\usepackage{ulem}
\usepackage{dsfont}
\usepackage{siunitx}
\DeclareSIUnit\gauss{G}
\usepackage{bibunits}
\usepackage{import}
\usepackage{braket}
\usepackage{nicefrac}
\usepackage{comment}
\usepackage{bm}
\usepackage{bbm}
\usepackage[english]{babel}
\usepackage{hyperref}

\usepackage{enumitem}

\usepackage[thinc]{esdiff}

\hypersetup{colorlinks=true,linkcolor=blue,citecolor=blue,breaklinks}

\newcommand{\hordimer}{\raisebox{-0.75 pt} {\includegraphics[scale=0.05]{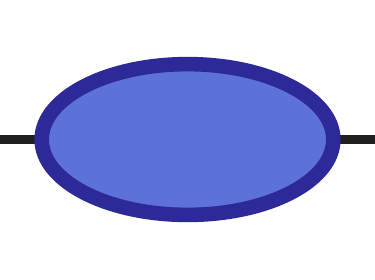}}}

\newcommand{\horDket}{| \raisebox{-0.75 pt} {\includegraphics[scale=0.05]{figures/symbols/Dimer_occ_updated.pdf}} \rangle}
\newcommand{\horDEket}{| \raisebox{-0.75 pt} {\includegraphics[scale=0.05]{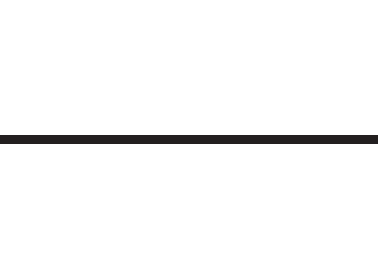}} \rangle}
\newcommand{\horDBket}{| {\includegraphics[scale=0.05]{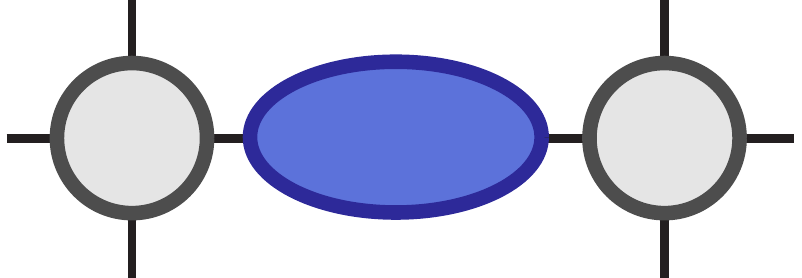}} \rangle}
\newcommand{\horMEket}{| {\includegraphics[scale=0.05]{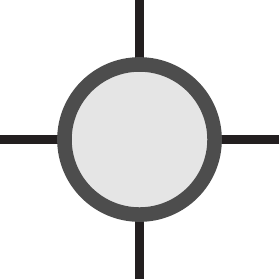}} \rangle}

\newcommand{\horMket}{| \raisebox{-0.75 pt} {\includegraphics[scale=0.05]{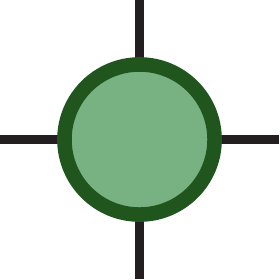}} \rangle}

\newcommand{\horDbra}{\langle \raisebox{-0.75 pt} {\includegraphics[scale=0.05]{figures/symbols/Dimer_occ_updated.pdf}} |}

\newcommand{\horMBbra}{\langle {\includegraphics[scale=0.05]{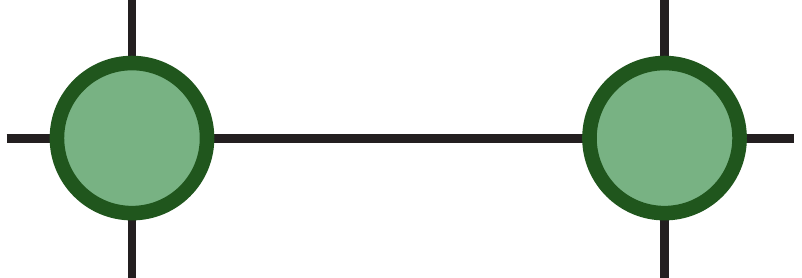}} |}
\newcommand{\horMbra}{\langle \raisebox{-0.75 pt}{\includegraphics[scale=0.05]{figures/symbols/Monomer_updated.pdf}} |}

\newcommand{\horUPket}{| \raisebox{-0.75 pt} {\includegraphics[scale=0.05]{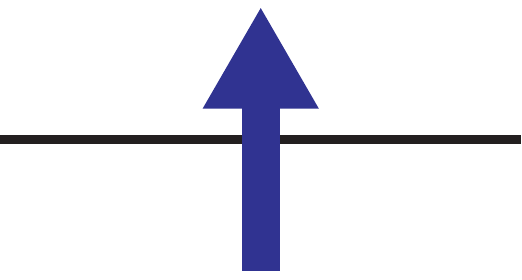}} \rangle}
\newcommand{\horDOWNket}{| \raisebox{-0.75 pt} {\includegraphics[scale=0.05]{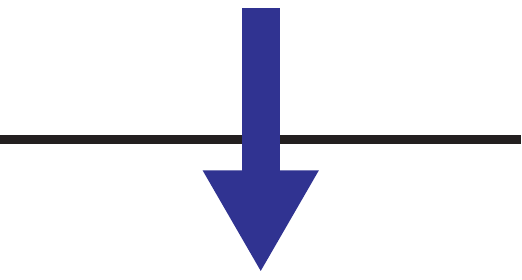}} \rangle}

\newcommand{\micro}[1]{\ensuremath{\mu\mathrm{#1}}}

\renewcommand{\micro}[1]{\ensuremath \mu\mathrm{#1}}
\newcommand{\um}{\ensuremath \mu\mathrm{m}}

\renewcommand{\vec}[1]{\ensuremath{\mathbf{#1}}}

\newcommand{\nm}{\ensuremath{\nu}}

\let\oldsfdefault\sfdefault
\usepackage[scaled=0.92]{helvet}
\renewcommand{\sfdefault}{\oldsfdefault}

\usepackage{layouts}

\usepackage{tikz}
\defaultbibliographystyle{apsrev4-2}

\begin{document}

\begin{bibunit}
\title{Dynamical preparation of $U(1)$ quantum spin liquids \\ in an analogue quantum simulator}

\author{Simon~Karch}\email{simon.karch@physik.uni-muenchen.de}
    \affiliation{Fakult\"{a}t f\"{u}r Physik, Ludwig-Maximilians-Universit\"{a}t, 80799 Munich, Germany}
    \affiliation{Max-Planck-Institut f\"{u}r Quantenoptik, 85748 Garching, Germany}
    \affiliation{Munich Center for Quantum Science and Technology (MCQST), 80799 Munich, Germany}
\author{Melissa~Will}
    \affiliation{Technical University of Munich, TUM School of Natural Sciences, Physics Department, 85748 Garching, Germany}
    \affiliation{Munich Center for Quantum Science and Technology (MCQST), 80799 Munich, Germany}
    \affiliation{Department of Physics, Harvard University, Cambridge, Massachusetts 02138, USA}
\author{Irene~Prieto~Rodriguez}
\author{Nikolas~Liebster}
    \affiliation{Fakult\"{a}t f\"{u}r Physik, Ludwig-Maximilians-Universit\"{a}t, 80799 Munich, Germany}
    \affiliation{Max-Planck-Institut f\"{u}r Quantenoptik, 85748 Garching, Germany}
    \affiliation{Munich Center for Quantum Science and Technology (MCQST), 80799 Munich, Germany}
\author{SeungJung~Huh}
    \affiliation{Fakult\"{a}t f\"{u}r Physik, Ludwig-Maximilians-Universit\"{a}t, 80799 Munich, Germany}
    \affiliation{Max-Planck-Institut f\"{u}r Quantenoptik, 85748 Garching, Germany}
    \affiliation{Munich Center for Quantum Science and Technology (MCQST), 80799 Munich, Germany}
    \affiliation{Department of Physics and Chemistry, Daegu Gyeongbuk Institute of Science and Technology (DGIST), Daegu 42988, Republic of Korea}
\author{Michael~Knap}
\author{Frank~Pollmann}
    \affiliation{Technical University of Munich, TUM School of Natural Sciences, Physics Department, 85748 Garching, Germany}
    \affiliation{Munich Center for Quantum Science and Technology (MCQST), 80799 Munich, Germany}
\author{Clemens~Kuhlenkamp}
    \affiliation{Department of Physics, Harvard University, Cambridge, Massachusetts 02138, USA}
\author{Immanuel~Bloch}
\author{Monika~Aidelsburger}\email{monika.aidelsburger@mpq.mpg.de}
    \affiliation{Fakult\"{a}t f\"{u}r Physik, Ludwig-Maximilians-Universit\"{a}t, 80799 Munich, Germany}
    \affiliation{Max-Planck-Institut f\"{u}r Quantenoptik, 85748 Garching, Germany}
    \affiliation{Munich Center for Quantum Science and Technology (MCQST), 80799 Munich, Germany}

\date{\today}


\begin{abstract}
Locally constrained gauge theories underpin our understanding of fundamental interactions in particle physics and the emergent behaviour of quantum materials. In strongly correlated systems, they can give rise to quantum spin liquids that lack conventional order and are defined by coherent superpositions of an extensive number of many-body configurations. Realising and probing such exotic states experimentally is an outstanding challenge both in solid-state and synthetic quantum systems, not least due to the difficulty of detecting the fragile coherences between many-body states. Here, we report a large-scale ($> 3{,}000$ sites) realisation of a two-dimensional $U(1)$ lattice gauge theory with ultracold atoms in a square optical superlattice and demonstrate non-equilibrium preparation of extended regions of $U(1)$ quantum spin liquids. We demonstrate Gauss’s law validity in a quench experiment, enabled by a new microscopy technique for detecting doubly occupied sites. We observe characteristic real-space correlations and momentum-space pinch points, hallmarks of the emergent $U(1)$ gauge structure. Using round-trip interferometric protocols, we directly observe large-scale coherence between many-body configurations, providing strong evidence for quantum spin liquid regions extending over $\sim100$ lattice sites. Our results establish non-equilibrium quantum simulation protocols as a powerful route for accessing and probing exotic, highly-entangled states beyond those hosted by the engineered Hamiltonian in thermal equilibrium.
\end{abstract}
\maketitle

\begin{figure*}[t!]
\includegraphics{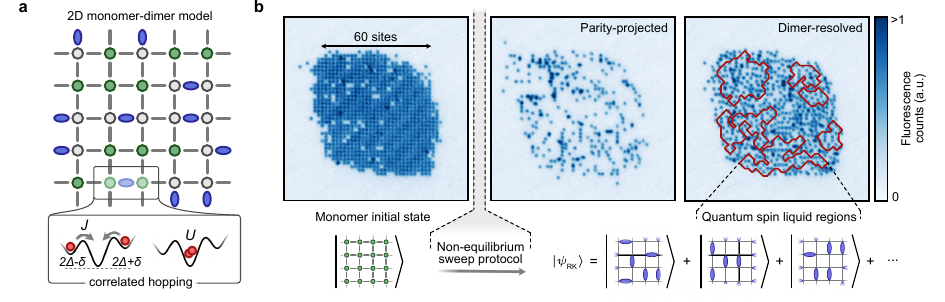}
     \caption{\textbf{Effective 2D monomer-dimer model and preparation of $U(1)$ quantum spin liquids.} \textbf{a,} Schematic of the 2D monomer-dimer model, where occupied vertices are monomers (green circles) and doublons on links are dimers (blue ellipses). This effective model arises from the 2D Bose-Hubbard model with tunnel coupling $J$, on-site interaction $U\gg J$, and staggered superlattice potential $\Delta \gg J$, restricting the dynamics to second-order correlated hopping processes $\ket{\cdots101\cdots} \leftrightarrow \ket{\cdots020\cdots}$ (inset). An additional linear tilt $\delta / J \sim 2$ suppresses competing second-order processes such as $\ket{\cdots100\cdots} \leftrightarrow \ket{\cdots001\cdots}$.
     \textbf{b,} Starting from the ordered monomer initial state (left, fluorescence image of more than $60 \times 60$ lattice sites), a semi-adiabatic sweep converts most monomers into dimers, preparing extended regions of $U(1)$ quantum spin liquid -- a massive superposition state of all allowed dimer coverings, as illustrated in the lower panel. Standard fluorescence imaging (centre, parity-projected) renders doublons as loss, so the prepared state appears largely empty. Dimer-resolved image (right, doublon-resolved) directly reveals patches of nearly perfect dimer coverings, outlined in red (see Fig.~\ref{fig:rk_experiment}a).}
     \label{fig:sketch}
\end{figure*}

Quantum states of matter at zero temperature are conventionally expected to be described by Landau's theory of spontaneous symmetry breaking, as in antiferromagnets or superfluids, or to be adiabatically connected to trivial product states, as in paramagnets or band insulators. Quantum spin liquids (QSLs) provide a striking counterexample of highly entangled states not captured by a local order parameter~\cite{savary_quantum_2017, wen_quantum_2007}. Instead, their physics is governed by local constraints and an emergent gauge structure. Anderson first proposed these as resonating valence bond states~\cite{anderson_resonating_1973}, in analogy to the delocalized electron configurations in a benzene ring. In this picture, spins on frustrated lattices, such as triangular antiferromagnets, form singlets that fluctuate between different configurations rather than freeze into a fixed order, even at zero temperature. Rokhsar and Kivelson~\cite{rokhsar_superconductivity_1988} later identified a paradigmatic realisation in the square-lattice quantum dimer model~\cite{moessner_short-ranged_2001,moessner_quantum_2008, fradkin_field_2013}, where dimers represent spin singlets and the ground state forms an equal-weight, equal-amplitude coherent superposition of all possible nearest-neighbour dimer coverings.

After decades of experimental efforts, conclusive evidence for QSLs in real materials remains elusive: candidate states are fragile, and direct signatures of large-scale many-body coherence are difficult to access. Despite these challenges, recent years have seen significant progress, with the identification of new candidate materials with possible QSL signatures~\cite{broholm_quantum_2020, breidenbach_identifying_2025, scheie_spectrum_2026}. Advances in analogue and digital quantum simulators offer an alternative route towards studying and manipulating these exotic states. They offer microscopic control and single-site read-out, thus enabling access to observables out of reach in solid-state experiments; however, local constraints do not arise naturally and must be engineered. Key advances have been made in implementing extended lattice gauge theories in one~\cite{bernien_probing_2017,surace_lattice_2020,yang_observation_2020,halimeh_cold-atom_2025,kebric_exploring_2025,datla_statistical_2026} and two dimensions~\cite{cochran_visualizing_2025,gonzalez-cuadra_observation_2025,gyawali_observation_2024,meth_simulating_2025,haghshenas_digital_2025,halimeh_quantum_2025}, establishing the foundation for exploring signatures of $\mathbb{Z}_2$ QSLs via nearest-neighbour blockade interactions in Rydberg atom arrays~\cite{semeghini_probing_2021} and digital state preparation in superconducting qubits~\cite{satzinger_realizing_2021}.

In this work, we realise extended $U(1)$ QSLs in a 2D \textit{monomer-dimer model} implemented in an analogue quantum simulator of neutral atoms in an optical lattice. The implementation is based on kinetically constrained dynamics in an optical superlattice, where correlated two-particle tunnelling governs the dynamics. Our approach is related to earlier 1D realisations~\cite{yang_observation_2020,zhou_thermalization_2022,wang_interrelated_2023} and proposed extensions to 2D~\cite{osborne_large-scale_2025}. The model is equivalent to a spin-1/2 quantum link model~\cite{horn_finite_1981,orland_lattice_1990,chandrasekharan_quantum_1997} of quantum electrodynamics (QED) in (2+1)D with dynamical matter~\cite{wiese_ultracold_2013,dalmonte_lattice_2016}. We first verify the validity of the effective model by probing constrained gauge dynamics in a quench experiment in a system with $> 3{,}000$ lattice sites and directly measuring Gauss's law. We then use non-equilibrium semi-adiabatic ramps to prepare $U(1)$ QSLs (Fig.~\ref{fig:sketch}b)~\cite{giudici_dynamical_2022,sahay_quantum_2023, gjonbalaj_shortcuts_2025}. This is achieved despite the fact that our Hamiltonian does not support this state as a ground state, and that $U(1)$ QSLs in 2D are typically critical, gapless, and only marginally stable~\cite{polyakov_1977}. We find the prepared QSL regions to be well described by a Rokhsar-Kivelson wavefunction, observing characteristic real-space dimer-dimer correlations and a hallmark of $U(1)$ gauge theory: momentum-space pinch points familiar from neutron scattering experiments on spin-ice materials~\cite{Isakov_2004,fennell_magnetic_2009, morris_dirac_2009}. Using round-trip interferometric protocols, we directly demonstrate the coherent many-body nature of the prepared state and quantify the spatial extent of the QSL regions to ${\sim}100$ lattice sites.

\begin{center}
\textbf{Experimental implementation}
\end{center}
\vspace{0.5em}

\begin{figure*}[t!]
\includegraphics{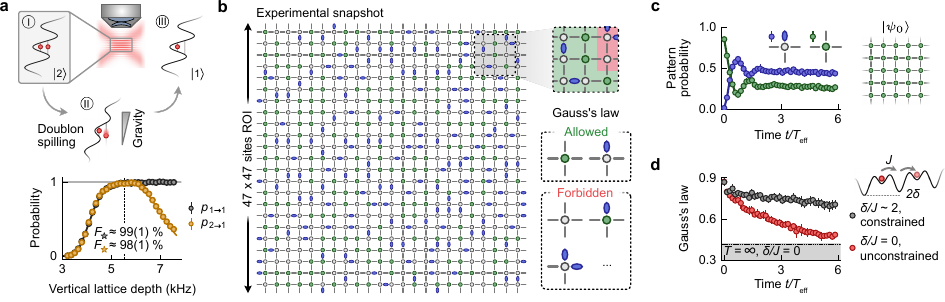}
     \caption{\textbf{Probing constrained gauge dynamics in the monomer-dimer model.}
     \textbf{a,} Schematic of the doublon-detection sequence, which converts doubly occupied sites into singly occupied sites prior to imaging. In the presence of a gravitational tilt, reducing the vertical lattice depth converts doublons into singlons with peak fidelity $98(1)\%$. The lower panel shows the conversion probability $p_{2\to 1}$ and the survival probability $p_{1\to 1}$ of singlons, as a function of vertical lattice depth. The error bars are smaller than the markers.
     \textbf{b,} Quantum gas microscope snapshot of a typical time-evolved state for a $47\times 47$-site ROI. The inset illustrates the two Gauss's-law-allowed vertex patterns together with the most common Gauss's-law-violating ones.
     \textbf{c,} Time evolution of the probabilities of the two Gauss's-law-allowed vertex patterns following a quench from the monomer initial state (initial fillings: $88(2)\%$ on vertices, $1.2(5)\%$ on links, and $0.2(2)\%$ on sites outside the effective Lieb lattice). The fraction of atoms on sites outside the effective Lieb lattice remains below $3\%$. Each data point is averaged over approximately eight experimental realisations; the error bars denote one standard deviation; where not visible, they are smaller than the marker.
     \textbf{d,} Fraction of Gauss's-law-satisfying vertices as a function of time following the quench, with (constrained gauge theory; grey) and without (unconstrained gauge theory; red) the diagonal tilt $\delta/J \sim 2$. The tilt suppresses direct hopping of monomers between vertices, slowing the decay of the Gauss's law validity. The dashed line delimiting the grey shaded region is a classical cellular automaton prediction for the unconstrained system at infinite temperature. The error bars denote one standard deviation; where not visible, they are smaller than the marker.
     }
     \label{fig:gauss-law}
\end{figure*}

Our experiments are performed using a caesium quantum gas microscope with tunable 2D optical superlattices~\cite{impertro_unsupervised_2023, karch_probing_2025}. The dynamics is governed by the 2D Bose-Hubbard Hamiltonian (Fig.~\ref{fig:sketch}a)
\begin{equation}
    \begin{aligned}
        \hat H &= - J\sum_{\langle i,j\rangle} \left( \hat a_{i}^{\dagger} \hat a_{j}^{\phantom{\ast}} + \mathrm{h.c.} \right) + \frac{U}{2} \sum_{i} \hat n_{i} (\hat n_{i}-1)\\ & + \sum_{i = (i_x, i_y)} \left[ \frac{(-1)^{i_x} + (-1)^{i_y}}{2} \Delta + (i_x+i_y)\delta \right]\hat n_{i},
    \end{aligned}
    \label{eq:bhm_hamiltonian}
\end{equation}
where $J/h\approx$ \SI{130}{Hz} is the tunnel coupling and $\delta\sim 2J$ a diagonal linear tilt. The dominant energy scales are the large staggered potential $\Delta/J \sim 12$, which suppresses single-particle tunnelling, and the on-site interaction $U/J \sim 24$. Near resonance $U \approx 2\Delta$, the dynamics is kinetically constrained and dominated by correlated two-particle tunnelling processes: two singlons (\textit{monomers}) on adjacent high-energy sites (\textit{vertices}) jointly hop to form a doublon (\textit{dimer}) on a low-energy site (\textit{link}), as shown in Fig.~\ref{fig:sketch}a. Starting from an initial state $\ket{\psi_0}$, where all vertices are occupied with singlons (all-monomer state), the constrained dynamics maps to an effective \textit{monomer-dimer} model on the Lieb lattice (see SI)
\begin{equation}
\hat{H}_{\text{eff}}= J_\mathrm{eff}\sum_{\mathrm{links}} \left( \horDBket \horMBbra + \text{h.c.} \right) + \sum_{\mathrm{vertices}} m \horMket \horMbra ,
    \label{eq:monomer-dimer}
\end{equation}
where $\horMket$, $\horDket $, $\horMEket $, $\horDEket$ denote a monomer, dimer, empty vertex and link respectively. The effective second-order coupling in the limit $\delta \ll U$ is $J_\mathrm{eff}\approx4\sqrt{2}J^2/U$, and the monomer cost -- the \textit{mass} -- is realized by tuning the correlated two-particle tunnelling away from resonance via the strength of the staggered potential $\Delta$, $m \approx \Delta - U/2$. The $U(1)$ lattice gauge theory in Eq.~(\ref{eq:monomer-dimer}) respects the following Gauss's law, defined around each vertex $V$
\begin{equation}
\hat{G}_V \equiv \horMket \horMbra+ \sum_{\mathrm{links\,at\,}V} \horDket \horDbra =1.
\end{equation}
The initial all-monomer state satisfies $\hat{G}_V\ket{\psi_0}= \ket{\psi_0}$ at every vertex, and the effective Hamiltonian (\ref{eq:monomer-dimer}) preserves this constraint under time evolution. 

In the language of QED in $(2{+}1)$D~\cite{fradkin_1990,fradkin_field_2013, wiese_ultracold_2013,dalmonte_lattice_2016}, monomers represent charges and link occupations encode the state of the electric field $\vec{E}$ ($\pm 1$ on empty links, and $\mp 3$ on links occupied by a dimer; see SI). The all-monomer state corresponds to a configuration of alternating positive and negative charges, and the pure dimer sector to the charge-free subspace of the lattice gauge theory.

\begin{center}
\textbf{Probing constrained gauge dynamics}
\end{center}
\vspace{0.5em}

To verify that the non-equilibrium dynamics is correctly described by the effective monomer-dimer Hamiltonian, we study the constrained $U(1)$ gauge dynamics after a quench within the Gauss's law sector. Directly probing Gauss's law requires simultaneous detection of monomers and dimers, which is not possible with standard quantum gas microscopy, as doubly occupied sites are parity-projected to empty sites during fluorescence imaging~\cite{bakr_quantum_2009,sherson_single-atom-resolved_2010,gross_quantum_2021}, but can be avoided by implementing doublon-resolved imaging~\cite{gross_quantum_2021, su_fast_2025}. Here we employ a new high-fidelity detection method that converts doubly occupied sites into singly occupied sites prior to imaging, revealing link occupations and hence directly enabling measurement of Gauss's law at every vertex and access to the dynamics of both matter and gauge degrees of freedom (Fig.~\ref{fig:gauss-law}a; see SI).

Starting from the excited all-monomer initial state, we quench to $m \approx 0$, where monomers resonantly couple to dimers on the links. The resulting time-evolved snapshots resolve both monomers and dimers across a $47 \times 47$-site region of interest (ROI) containing more than 500 vertices and 1,000 links, enabling a direct evaluation of Gauss's law by classifying the occupation pattern around every vertex (Fig.~\ref{fig:gauss-law}b). Two patterns satisfy Gauss's law: vertices occupied by monomers without any dimers on neighbouring links, and empty vertices with exactly one dimer attached. The most common violations are empty vertices with no dimers attached, double-dimer configurations, and monomer-dimer configurations.

The dynamics of matter and gauge degrees of freedom (Fig.~\ref{fig:gauss-law}c) equilibrates locally within $\sim2T_{\mathrm{eff}}$, where $T_{\mathrm{eff}} = \hbar / J_\mathrm{eff} =\, $\SI{5.13(7)}{ms} is the effective tunnelling time, while the fraction of Gauss's-law-satisfying vertices decays only slowly (Fig.~\ref{fig:gauss-law}d).

To quantify the effective dimension of the Hilbert space explored during the dynamics, we evaluate the subsystem return probability~\cite{karch_probing_2025} from snapshots in the regime $t>4T_\mathrm{eff}$ and find $ \mathrm{dim}\,\mathcal{H}_{\mathrm{eff}} \approx 2.9^A$, where $A$ is the number of vertices in the subsystem. In the absence of the tilt ($\delta/J = 0$), direct monomer hopping between vertices is allowed and the fraction of Gauss's-law-satisfying vertices decays significantly faster (Fig.~\ref{fig:gauss-law}d), giving $\mathrm{dim}\,\mathcal{H}_{\mathrm{eff}} \approx 4.4^A$. This increase reflects the breakdown of the effective gauge structure.

\begin{center}
\textbf{Non-equilibrium preparation of QSLs}
\end{center}
\vspace{0.5em}

\begin{figure}[t!]
\includegraphics{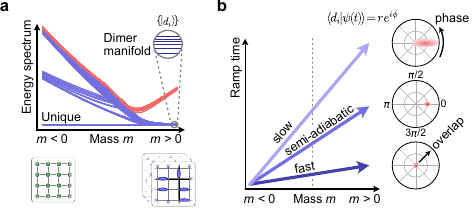}
     \caption{\textbf{Non-equilibrium preparation of $U(1)$ quantum spin liquids.}
     \textbf{a,} Schematic energy spectrum as a function of monomer mass $m$. For large negative monomer mass $m$, the ground state is unique, with all vertices occupied by monomers and no dimers (left inset). For large positive $m$, the lowest-energy states form a nearly degenerate manifold of dimer coverings $\{ \ket{d_i} \}$ (right inset), with small splittings arising from higher-order terms such as plaquette interactions. Blue curves are the states connected to the dimer manifold, and red curves represent excited states.
     \textbf{b,} Schematic of the dynamical mass sweep from $m<0$ to $m>0$, starting from the monomer initial state. The polar plots sketch the overlaps $\langle d_i |\psi(t)\rangle  = re^{i\phi}$ with individual dimer coverings $\ket{d_i}$ for three sweep rates. A fast ramp (dark blue) results in poor overlap with the dimer manifold ($r \approx 0$). A slow ramp (light blue) produces unequal coefficients $re^{i\phi}$, preferring the lowest-energy eigenstates within the dimer manifold, which have dimer ordering. A semi-adiabatic ramp (medium blue) populates all dimer coverings with approximately equal amplitudes and phases, realising the $U(1)$ QSL state.}
     \label{fig:rk_theory}
\end{figure}

\begin{figure*}[t!]
\includegraphics{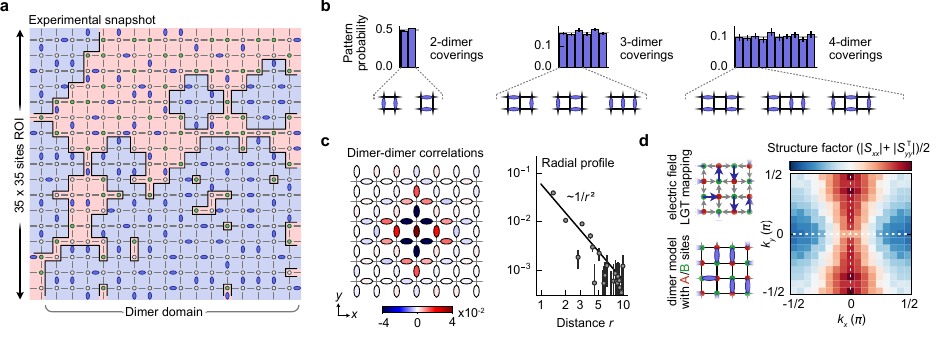}
     \caption{\textbf{Experimental preparation of $U(1)$ QSLs.}
     \textbf{a,} Representative experimental snapshot after the forward ramp, with a $35 \times 35$-site ROI indicated. Blue shading marks domains of near-perfect dimer covering consistent with the RK wavefunction; red shading marks regions containing monomers and Gauss's law defects. The probability of a vertex occupied by a single monomer with no attached dimers is $14.9(1)\%$, and the probability of a single dimer attached to a vertex is $68.4(2)\%$, giving a Gauss's law validity of $83.2(2)\%$. The dominant violation is due to empty sites [$9.5(1)\%$], followed by monomer-plus-dimer [$4.3(1)\%$] and double-dimer [$2.7(1)\%$] configurations.
     \textbf{b,} Pattern probabilities for subsystem configurations containing two, three, and four dimers. The approximately equal probabilities of all distinct dimer coverings are consistent with the RK wavefunction. Results are obtained from 431 experimental snapshots, error bars denote one standard deviation from bootstrapping.
     \textbf{c,} Spatial map of connected dimer-dimer correlations (left) and corresponding radial profile (right), as a function of distance $r=\sqrt{x^2+y^2}$ from a central vertical link (assuming spatial inversion symmetry). We have excluded links attached to Gauss's law-violating vertices in the analysis. Error bars denote one standard deviation from bootstrapping. The solid line is a $1/r^2$ guide to the eye.
     \textbf{d,} $U(1)$ QSL represented as electric fields in a (2+1)D LGT, and the corresponding structure factor in momentum space. Empty links carry an alternating electric field of amplitude $\pm1$ pointing outward/inward on A/B sites. A dimer on a link multiplies the field by $-3$, mapping Gauss's law onto the ice rule. The structure factor $S_{\alpha\alpha}(\mathbf{k})$ ($\alpha=x,y$) is calculated using the Fourier transform of the measured electric-field correlations and exhibits a characteristic pinch point at $\mathbf{k}=(0,0)$. The plotted quantity $\frac{1}{2}\left( |S_{xx}|+|S_{yy}^T| \right)$ averages $|S_{xx}|$ and $|S_{yy}|$ after a $90^\circ$ rotation, and is further symmetrised using the mirror symmetries (white dashed lines).
     }
     \label{fig:rk_experiment}
\end{figure*}

\begin{figure*}[t!]
\includegraphics{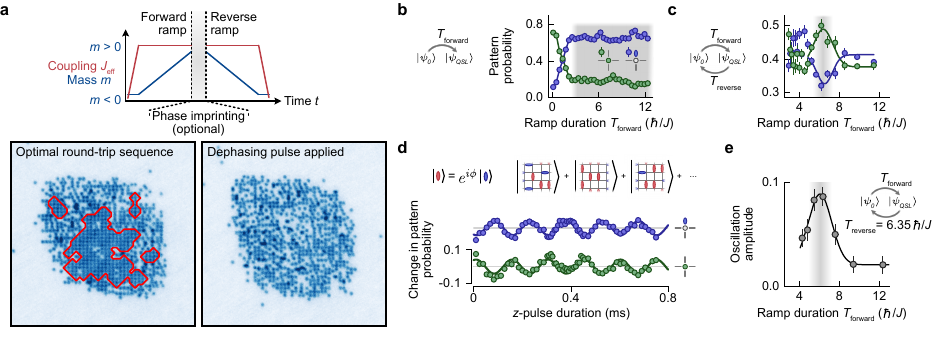}
     \caption{\textbf{Round-trip probe for many-body coherence.}
     \textbf{a,} Schematic of the round-trip interferometric protocol (top). The coupling $J_{\mathrm{eff}}$ (red) is switched on in \SI{1.8}{ms} and then, the mass is ramped forward into the QSL state. Phase imprinting is optionally applied and the forward ramp is then reversed. Selected fluorescence images (bottom) show the optimal round-trip, in which large initial monomer state domains emerge (left, outlines in red), and a round-trip with a dephasing pulse applied before the reverse ramp that suppresses the appearance of the patches (right).
     \textbf{b,} Detection probabilities of the two Gauss's-law-allowed vertex patterns (single monomer, green; single dimer, blue) as a function of ramp duration $T_{\mathrm{forward}}$, measured directly after the forward ramp from the monomer initial state ($\ket{\psi_0}$) into the QSL ($\ket{\psi_{\mathrm{QSL}}}$). For $T_{\mathrm{forward}} \gtrsim 2.5\hbar/J$ the pattern probabilities reach an approximate plateau (grey shaded region). The error bars denote the standard error of the mean (s.e.m.); where not visible, they are smaller than the marker.
     \textbf{c,} Post round-trip detection probabilities of the same two patterns as a function of ramp duration ($T_{\mathrm{forward}}$ = $T_{\mathrm{reverse}}$). The solid line is a Gaussian fit as a guide to the eye; the grey shaded region indicates the optimal ramp duration, centred at $T_{\mathrm{forward}}\approx6.4(1)\,\hbar/J$. The error bars denote the s.e.m.; where not visible, they are smaller than the markers.
     \textbf{d,} An abrupt increase in the staggered potential depth along one lattice axis before the reverse ramp imprints a phase between the vertical and horizontal dimers [$T_\mathrm{forward} = T_\mathrm{reverse} = 6.35(3)\,\hbar/J$]. The two allowed vertex patterns oscillate out-of-phase at \SI{6.74(4)}{kHz}. The error bars denote the s.e.m.; where not visible, are smaller than the markers.
     \textbf{e,} Peak-to-peak oscillation amplitude (mean over monomer and dimer vertex patterns) extracted as a function of $T_\mathrm{forward}$, with $T_\mathrm{reverse} = 6.35(3)\,\hbar/J$. The shaded region is centred at $T_{\mathrm{forward}}\approx6.0(1)\,\hbar/J$. Error bars are the fit errors.
     }
     \label{fig:coherences}
\end{figure*}

Having verified that the engineered potential energy landscape correctly implements the monomer-dimer model with Gauss's law constraints, we now leverage the gauge theory implementation to study the non-equilibrium dynamical preparation of $U(1)$ QSLs.

In the limit of large negative mass $m < 0$, dimers are energetically costly and the monomer state is the unique ground state. For large positive mass $m > 0$, it becomes favourable to pair all monomers into dimers, but there is no unique way to do so. The resulting dimer manifold contains exponentially many nearly degenerate states $\ket{d_i}$ corresponding to all possible nearest-neighbour dimer coverings of the square lattice. The degeneracy is lifted by higher-order terms not captured by the effective second-order Hamiltonian, such as the (eighth-order) plaquette coupling $\sim\,J^8/U^7$, which splits coverings with distinct numbers of flippable plaquettes, that is maximal for columnar and zero for staggered configurations (Fig.~\ref{fig:rk_theory}a). A slow ramp preferentially populates the lower-energy coverings (Fig.~\ref{fig:rk_theory}b). A very fast ramp, on the other hand, produces a large number of unpaired monomer defects and poor overlap with the dimer manifold.

Between these two limits lies the \textit{semi-adiabatic} regime~\cite{semeghini_probing_2021,giudici_dynamical_2022, sahay_quantum_2023, gjonbalaj_shortcuts_2025}. Such a ramp is slow enough to be adiabatic with respect to monomer excitations, which cost energy $2|m|$, but fast enough to be sudden with respect to the small splittings within the dimer manifold. The non-equilibrium preparation is aided by the smallness of the plaquette coupling, which would otherwise lift the degeneracy of the dimer coverings. The result is extended regions of near equal-amplitude, equal-phase superposition of all dimer coverings $\{\ket{d_i}\}$, described by the RK wavefunction~\cite{rokhsar_superconductivity_1988},
\begin{equation}
    \ket{\Psi_{\mathrm{RK}}} = \frac{1}{\sqrt{N}} \sum_i \ket{d_i},
\end{equation}
where $N$ denotes the number of nearly-degenerate dimer coverings.

\begin{center}
\textbf{Experimental preparation of $U(1)$ QSLs}
\end{center}
\vspace{0.5em}

We implement the semi-adiabatic sweep by linearly ramping the mass $2m$ from $-3.8(1)J_\mathrm{eff}$ to $+3.6(1)J_\mathrm{eff}$ via the staggered potential ($\Delta_x = \Delta_y$) over a duration $T_\mathrm{forward} = 6.35(3)\,\hbar/J$. The ramp duration is determined by maximising the monomer filling after a round-trip experiment (see below). The prepared state fulfils Gauss's law on $83.2(2)\%$ of the vertices, with the dominant violation arising from empty vertices in the initial monomer state.

Experimental snapshots confirm the structure of the state expected from the semi-adiabatic protocol (Fig.~\ref{fig:rk_experiment}a): we observe that regions of near-perfect dimer covering are interspersed with regions of unpaired monomers. The detection probabilities of distinct two-, three-, and four-dimer subsystem configurations are approximately equal (Fig.~\ref{fig:rk_experiment}b), consistent with the notion of the RK wavefunction being an equal-amplitude superposition of dimer coverings, and in contrast to an ordered state, such as columnar phases that would have dimers aligned along $x$- or $y$-axes~\cite{rokhsar_superconductivity_1988}, which we can prepare by breaking the symmetry between the two superlattice axes.

Next, we evaluate connected dimer-dimer correlations between a vertical link at position $\vec r_i$ and other links at positions $\vec r_j$, $\langle \hordimer_{\vec{r}_i} \hordimer_{\vec{r}_j}\rangle - \langle \hordimer_{\vec{r}_i} \rangle \langle \hordimer_{\vec{r}_j}\rangle$. The ideal Rokhsar-Kivelson $U(1)$ QSL is critical, and the dimer-dimer correlations are predicted to decay algebraically as $1/|\vec r_i - \vec r_j|^2$. We observe a radial profile consistent with this power law (Fig.~\ref{fig:rk_experiment}c).

Using the electric-field mapping in the language of lattice QED introduced above, we compute the electric-field structure factor from the snapshots, $S_{\alpha\alpha}(\mathbf{k}) = \langle \vec{E}_\alpha(\mathbf{k}) \vec{E}_\alpha(-\mathbf{k})\rangle_c$ ($\alpha = x, y$; subscript $c$ denotes the connected correlator). We observe a characteristic pinch point at $\vec{k} = \vec 0$: a bow-tie shaped feature that is a direct consequence of Gauss's law in momentum space $\vec\nabla \cdot \vec{E}(\vec{r}) = 0 \leftrightarrow \vec{k} \cdot \vec{E}(\vec{k}) = 0$, which forces the electric field to be transverse (Fig.~\ref{fig:rk_experiment}d). This directly connects our experiment to the physics of spin-ice materials, where the ice rule plays the role of Gauss's law~\cite{bramwell_spin_2001, fradkin_field_2013}, and pinch points are observed in neutron scattering experiments~\cite{fennell_magnetic_2009, morris_dirac_2009}.

\begin{center}
\textbf{Large-scale many-body coherence}
\end{center}
\vspace{0.5em}

\begin{figure}[t!]
\includegraphics{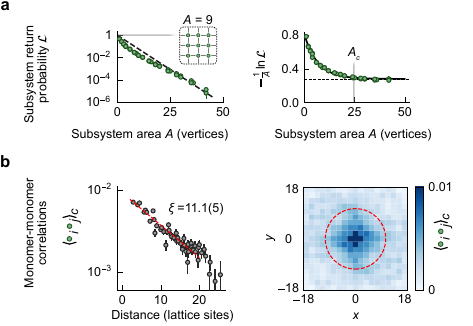}
     \caption{\textbf{Length scale of the QSL regions.}
     \textbf{a,} Subsystem return probability $\mathcal{L}$ evaluated from round-trip snapshots [$T_\mathrm{forward} = T_\mathrm{reverse} = 6.35(3)\,\hbar/J$] as a function of the subsystem area $A$, for rectangular subsystems such as the one illustrated in the inset (left). The normalised subsystem return probability $-\frac{1}{A}\ln\,\mathcal{L}$ (right) saturates to a constant (dashed line) for subsystems larger than the crossover area $A_c$, extracted by fitting an exponential with a constant offset (within $5\%$ of the constant value). Error bars are estimated using standard error of proportion; where not visible, are smaller than the markers.
     \textbf{b,} Connected monomer-monomer correlations from the round-trip snapshots. Left: radial average with exponential fit (red dashed line), giving correlation length $\xi = 11.1(5)$ lattice sites. Right: two-dimensional monomer-monomer correlation map; red dashed circle has radius $\xi$. Error bars denote one standard deviation from bootstrapping.
     }
     \label{fig:qsl_lengthscale}
\end{figure}

The dimer-dimer correlations, pinch points, and equal-amplitude histograms above are consistent with a $U(1)$ QSL, but equally compatible with an incoherent mixture of dimer coverings, analogous to thermal mixtures realised in classical spin ice materials~\cite{harris_geometrical_1997, ramirez_zero-point_1999, bramwell_spin_2001, fennell_magnetic_2009, morris_dirac_2009, pomaranski_absence_2013, gingras_spin_2021}. Demonstrating experimentally the massive many-body phase coherence, the defining property of any QSL, remains an open challenge. We address this question with a round-trip interferometric protocol (Fig.~\ref{fig:coherences}a): a forward semi-adiabatic ramp from $m<0$ to $m>0$ prepares the state explored in Fig.~\ref{fig:rk_experiment}, and a time-reversed ramp attempts to map the state back onto the monomer initial state. High return probability to the initial monomer state (i.e., high round-trip fidelity) requires the forward semi-adiabatic ramp to have produced a coherent superposition state of dimer coverings, while an incoherent mixture of dimer coverings would lead to a lower round-trip fidelity (see SI for supporting numerics). The monomer filling after the round-trip is therefore a proxy for the many-body phase coherence between the dimer coverings.

The ramp duration and endpoints are optimised by maximising the monomer filling in a symmetric round-trip experiment ($T_\mathrm{forward} = T_\mathrm{reverse}$). At the optimum [$T_\mathrm{forward} = T_\mathrm{reverse} = 6.4(1)\,\hbar/J$], large patches of the initial monomer state emerge in the snapshots (Fig.~\ref{fig:coherences}a). These patches largely vanish when a short dephasing pulse -- implemented using an incommensurate periodic diagonal potential -- is inserted before the reverse ramp, confirming that phase coherence is a prerequisite for high round-trip fidelity. Probing the state directly after the forward ramp as a function of $T_\mathrm{forward}$ (Fig.~\ref{fig:coherences}b), we find that once $T_{\mathrm{forward}} \gtrsim 2.5\,\hbar/J$ the prepared state is strikingly insensitive to the ramp duration. In contrast, in the round-trip experiment ($T_\mathrm{forward} = T_\mathrm{reverse}$), we observe a strong dependence of the monomer filling on the ramp duration (Fig.~\ref{fig:coherences}c), finding a sharp peak at $\approx6.4(1)\,\hbar/J$. Fixing the reverse ramp duration at the optimum and scanning $T_{\mathrm{forward}}$, we observe a similarly sharp peak (see SI). This reflects the high phase-sensitivity of the round-trip interferometer, which is enhanced by the dynamical phases that are picked up during the ramp.

The high sensitivity of the round-trip protocol is the key ingredient for using it to manipulate and detect the relative phases between distinct dimer coverings. We manipulate their relative phases by introducing an energy difference between horizontal and vertical dimers for a variable amount of time before the reverse ramp (Fig.~\ref{fig:coherences}d). This is achieved via an abrupt jump of the staggered potential by an amount $\delta\Delta = \Delta_y - \Delta_x$. Because each dimer covering $\ket{d_i}$ accumulates a total phase $\Phi_i$ determined by the number of horizontal versus vertical dimers, the global phase coherence of the state before the reverse ramp is scrambled unless \mbox{$\Phi_i = 2\pi n\,(n\in\mathbb{Z})$}, and the overlap $\left| \bra{\psi_\mathrm{RK}}\left(\sum_ie^{i\Phi_i}\ket{d_i}\bra{d_i}\right)\ket{\psi_\mathrm{RK}} \right|^2$ vanishes. Correspondingly, we observe periodic out-of-phase oscillations in the detection probability of both Gauss's law allowed patterns. In contrast, no oscillations are expected for an incoherent mixture of dimer coverings or for columnar states, which are prepared by applying a chemical potential bias to horizontal or vertical links during the forward ramp (see SI). Scanning $T_\mathrm{forward}$ at fixed $T_\mathrm{reverse}$ (Fig.~\ref{fig:coherences}e), we find that the oscillation amplitude follows the same trend as the post round-trip monomer filling, peaking sharply at $T_\mathrm{forward}\approx6.0(1)\,\hbar/J$. Together these measurements provide strong evidence for large-scale coherences between the many-body configurations.

\begin{center}
\textbf{Length-scale of the QSL regions}
\end{center}
\vspace{0.5em}

The emergence of large monomer domains in the round-trip snapshots (Fig.~\ref{fig:coherences}a), together with large domains of near-perfect dimer covering after the forward ramp (Fig.~\ref{fig:rk_experiment}a), suggests that the non-equilibrium semi-adiabatic ramp prepares finite-size regions of phase-coherent QSLs. Since phase coherence is a prerequisite for high round-trip fidelity, the round-trip snapshots provide a direct handle on the size of these regions.

To quantify the length scale of the monomer patches after the round-trip, we evaluate the subsystem return probability $\mathcal{L}_A $, i.e., the probability for a subsystem with $A$ vertices to be found in the initial all-monomer state configuration~\cite{karch_probing_2025}. As the subsystem size grows from $A = 1$ ($\mathcal{L}_1$ is equivalent to the probability of detecting a vertex occupied by a monomer with no dimers attached), $\mathcal{L}_A$ progressively captures all genuine $n$-point correlations up to order $n = A$. Once $A$ exceeds the area $A_c$ that contains all relevant many-body correlations, the subsystem return probability crosses over to the trivial exponential scaling, $\mathcal{L}_A \sim e^{-\alpha A}$, where $\alpha$ is a constant.

In Fig.~\ref{fig:qsl_lengthscale}a we plot the subsystem return probability and the normalised quantity $-\frac{1}{A}\ln\,\mathcal{L}$ as a function of subsystem area. We find that $-\frac{1}{A}\ln\,\mathcal{L}$ initially drops with increasing subsystem size and saturates to a constant at $A_c = 25$ vertices, corresponding to a disc of diameter $\approx 11$ lattice sites. The connected monomer-monomer correlations (Fig.~\ref{fig:qsl_lengthscale}b) are radially symmetric and decay exponentially with a correlation length $\xi = 11.1(5)$ lattice sites, in quantitative agreement with $A_c$, both covering an area of ${\sim}100$ lattice sites. As an independent measure, we analyse the probability of detecting subsystems with near-perfect dimer coverings in snapshots after the forward ramp (Fig.~\ref{fig:rk_experiment}; see SI). We find a characteristic area $A_c = 28$ vertices, in good agreement with the $A_c = 25$ vertices extracted from the round-trip snapshots. Together, these results strongly suggest that the monomer patches observed in the round-trip experiment appear precisely in regions where the forward ramp has prepared a $U(1)$ QSL, providing a quantitative estimate of the size of the QSL regions.

\begin{center}
\textbf{Discussion and outlook}
\end{center}
\vspace{0.5em}

In this work, we have demonstrated a $(2{+}1)$D $U(1)$ lattice gauge theory realised with ultracold atoms, using it to dynamically prepare regions of $U(1)$ QSL spanning $\sim 100$ lattice sites, well described by the Rokhsar-Kivelson wavefunction. We observe characteristic real-space correlations, momentum-space pinch points, and provide direct evidence for large-scale many-body coherence between the many-body configurations using round-trip interferometric protocols. This is achieved despite the fact that $U(1)$ QSLs in two dimensions are unstable towards perturbations~\cite{polyakov_1977} and that our Hamiltonian does not support the Rokhsar-Kivelson wavefunction as its ground state. It remains an open question which other exotic states can be dynamically prepared, and whether the overlap with the RK QSL can be improved using counterdiabatic or shortcut-to-adiabaticity protocols~\cite{gjonbalaj_shortcuts_2025}. Interesting future directions further include studying the dynamics of excitations, extensions to three dimensions, where emergent photon excitations~\cite{fradkin_field_2013, gao_neutron_2025} could be probed, or engineering plaquette interactions in a hybrid digital-analogue implementation in our platform~\cite{impertro_local_2024}.

\textit{Note:} During the preparation of the manuscript we became aware of related works exploring $U(1)$ spin liquids in Rydberg atom arrays~\cite{geim_engineering_2026, bornet_dirac_2026}.

\vspace{2em}
\begin{center}
\textbf{ACKNOWLEDGEMENTS}
\end{center}
\vspace{0.5em}

The authors would like to acknowledge insightful discussions with C. Chin, F. Desrochers, S. F\"olling, M. Lukin and his team, P. Preiss, and T. Zache. This work was supported by the Deutsche Forschungsgemeinschaft (DFG, German Research Foundation) via Research Unit FOR5522 under project number 499180199  and under Germany’s Excellence Strategy – EXC-2111 – 390814868 and from the German Federal Ministry of Education and Research via the funding program quantum technologies – from basic research to market (contract number 13N15895 FermiQP). This publication has further received funding under Horizon Europe programme HORIZON-CL4-2022-QUANTUM-02-SGA via the project 101113690 (PASQuanS2.1), the European Union (grant agreement No 101169765), as well as the Munich Quantum Valley, which is supported by the Bavarian state government with funds from the Hightech Agenda Bayern Plus. M.W. was supported by a scholarship of the German Academic Exchange Service (DAAD). I.P.R. received funding from the International Max Planck Research School (IMPRS) for Quantum Science and Technology. S.H. was supported by the education and training program of the Quantum Information Research Support Center, funded through the National research foundation of Korea (NRF) by the Ministry of science and ICT (MSIT) of the Korean government (No.2021M3H3A1036573). C.K. acknowledges funding from the Swiss National Science Foundation (Postdoc.Mobility Grant No. 217884).

\vspace{2em}
\begin{center}
\textbf{REFERENCES}
\end{center}
\vspace{0.5em}

\putbib[manuscript]
\end{bibunit}

\clearpage
\begin{bibunit}
\setcounter{section}{0}
\setcounter{equation}{0}
\setcounter{figure}{0}
\setcounter{table}{0}
\renewcommand{\theequation}{S\arabic{equation}}
\renewcommand{\theHequation}{S\arabic{equation}}
\renewcommand{\thefigure}{S\arabic{figure}}
\renewcommand{\theHfigure}{S\arabic{figure}}
\renewcommand{\thetable}{S\arabic{table}}
\renewcommand{\theHtable}{S\arabic{table}}
\setcounter{page}{1}

\title{Supplementary Information for: \\ Dynamical preparation of $U(1)$ quantum spin liquids \\ in an analogue quantum simulator}

\maketitle
\tableofcontents


\section{Experimental details}

\subsection{Experimental setup}

Our experiments begin with a quasi-pure Bose-Einstein condensate of $\sim 20,000-30,000$ $^{133}\textrm{Cs}$ atoms prepared in a crossed dipole trap, which is subsequently loaded into a single plane of a shallow-angle vertical lattice with $\SI{8}{\micro\m}$ spacing. The atoms are radially confined in a deep repulsive box potential generated using a digital micromirror device (DMD) illuminated by an incoherent light source from a multimode diode with central wavelength $\sim\SI{525}\nm$ (Wavespectrum WSLX-525-005-M-F22, power output $>\SI{4.5}{W}$). We first evaporate the atoms along the vertical direction and subsequently prepare Mott insulating states by ramping up the horizontal lattices (see Section~
\ref{sec:experimental_sequence}). The superlattice potential along each lattice axis is described by 
\begin{align}
    V(x) = V_\mathrm{s} \cos^2\left(\pi x/a_\mathrm{s}\right) + V_\mathrm{l} \cos^2\left(\pi x /a_\mathrm{l} + \phi/2\right),
\end{align}
where $V_\mathrm{s(l)}$ is the depth of the short(long)-period lattice, $\phi$ is the superlattice phase ($\phi=\pi/2$ corresponds to the staggered potential used in our experiments), and $a_s = \SI{383.5}{nm}, a_l = \SI{767}{nm}$ are the lattice spacings.

This product initial state is then transferred from the shallow-angle vertical lattice to a steep-angle vertical lattice formed by interfering two $\SI{1064}{nm}$ beams at an angle of $60^{\circ}$ (spacing $\approx$\SI{1}{\um}). The phase of this lattice is controlled by a ring piezo actuator that adjusts the path difference between the two interfering beams, and is set to ensure the atoms are loaded into a single plane of the steep-angle vertical lattice. We then perform an additional vertical spilling step in the deep horizontal two-dimensional lattice, converting remaining doublons into singlons, as described in more detail in Section~\ref{sec:doublon_spilling}. This step is essential for preparing high-quality initial states over large system sizes. When preparing large initial states, the residual harmonic confinement of the lattices becomes non-negligible and a doublon shell typically forms in the centre of the system. The spilling step removes these excess doublons.

Following the experimental sequence described in Section~\ref{sec:experimental_sequence}, we rapidly freeze the atoms in a deep $60\,E_r^{\mathrm{s}}$ two-dimensional lattice to project the wavefunction onto the Fock basis. We subsequently perform fluorescence imaging with a 0.8-NA high-resolution objective (Special Optics). Further details about the experimental setup can be found in Refs.~\cite{impertro_unsupervised_2023,wienand_emergence_2024,impertro_local_2024,impertro_strongly_2025,karch_probing_2025}.

\subsection{Experimental sequence}\label{sec:experimental_sequence}

\begin{figure*}[ht!]
\includegraphics{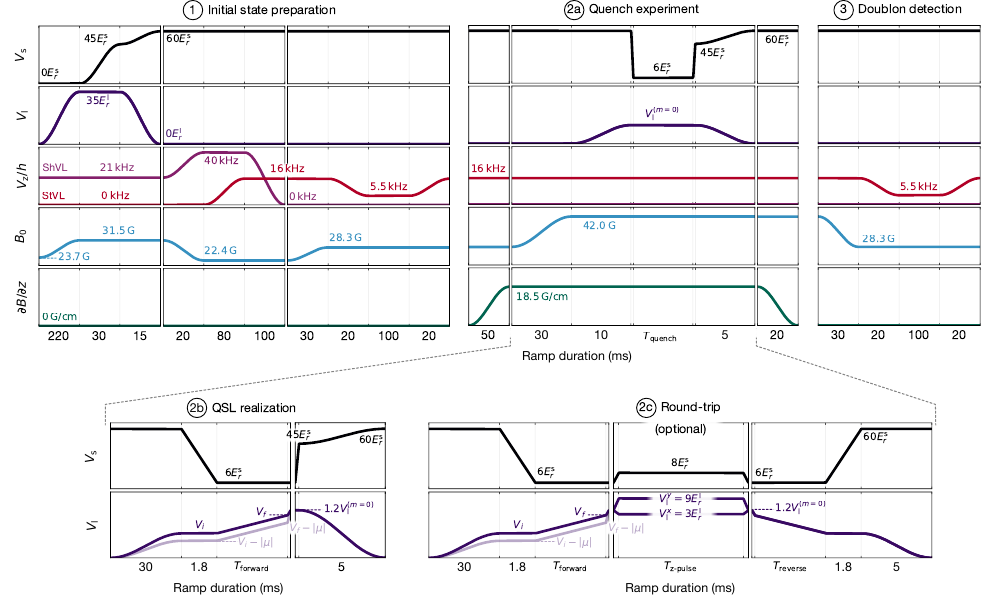}
     \caption{\textbf{Experimental sequence.} Programmed ramps of the short ($V_{\mathrm{s}}$) and long ($V_{\mathrm{l}}$) horizontal lattices, shallow- (ShVL) and steep-angle (StVL) vertical lattices ($V_z$), magnetic offset field ($B_0$), and magnetic field gradient in the $z$-direction ($\partial B/\partial z$, produced by a quadrupole field). \textbf{1,} Initial state preparation.
     \textbf{2,} Non-equilibrium dynamics: quench experiment (a), preparation of QSL states (b), and round-trip protocol with optional phase imprint (c). Long lattice ramps used for preparing columnar states are indicated in light purple.
     \textbf{3,} Doublon-to-singlon conversion sequence. The $x$- and $y$-axes are not to scale. Unlabelled ramps have a duration of \SI{0.15}{ms}.}
     \label{fig:suppmat:exp_sequence}
 \end{figure*}

The experimental sequence is depicted in Fig.~\ref{fig:suppmat:exp_sequence} and described in the following:\\

\textbf{1) Initial state preparation.}

The initial state is prepared by first ramping the long lattices to $35\,E_r^{\mathrm{l}}$ over $220\,\mathrm{ms}$ (at $\phi\sim-0.05\pi$), and then increasing the short lattice depth to $45\,E_r^{\mathrm{s}}$ over $20\,{\mathrm{ms}}$, localizing the atoms at the lowest-energy site of the $2\times2$ unit cell. Here, $E_r^{\mathrm{s,l}}=h^2/(8ma_{\mathrm{s,l}}^2)$ is the recoil energy of the short and long horizontal lattices, respectively. During the long lattice ramp, we also change the offset field from $23.7\,\mathrm{G}$ to $31.5\,\mathrm{G}$, increasing the scattering length from $a\sim330\,a_0$ to $a\sim580\,a_0$, where $a_0$ is the Bohr radius. We then increase the short lattices to $60\,E_r^{\mathrm{s}}$ while simultaneously turning off the long lattices.  At this point, lattice sites corresponding to vertices in the monomer-dimer model are occupied by singlons, with some
residual doublons in the centre of the system.

The atoms are then transferred from the shallow-angle vertical lattice (ShVL) to the steep-angle vertical lattice (StVL) which has lower potential disorder and a larger vertical trapping frequency $\omega_z$~\cite{impertro_strongly_2025}. To this end, the offset field is ramped to $22\,\mathrm{G}$ over $20\,{\mathrm{ms}}$, close to the three-body loss minimum for caesium, reducing the spatial extent of the doublon wavepackets. Simultaneously, the ShVL depth is increased from $21\,\mathrm{kHz}$ to $40\,\mathrm{kHz}$, compressing the atoms vertically. The StVL is then ramped up over $80\,\mathrm{ms}$ to $16\,{\mathrm{kHz}}$, and the ShVL is removed over $100\,\mathrm{ms}$. During this handover, the superlattice phase is changed to $\phi = \pi/2$, corresponding to the staggered configuration.

Finally, we perform an additional vertical spilling step, converting residual doublons into singlons, as described in more detail in Section~\ref{sec:doublon_spilling}, thereby completing the preparation of the monomer initial state. The offset field is ramped to $28.3\,\mathrm{G}$ over $30\,\mathrm{ms}$, setting the scattering length to $a\sim490\,a_0$. The StVL depth is then reduced from $16\,\mathrm{kHz}$ to $5.5\,\mathrm{kHz}$ over $20\,\mathrm{ms}$, held for $100\,\mathrm{ms}$, and subsequently ramped back up.\\

\textbf{2) Non-equilibrium dynamics.}

The diagonal tilt is implemented by applying a quadrupole field using a pair of coils along the $z$-direction and rotating the offset field into the diagonal direction in the horizontal plane. During this rotation, the offset field magnitude is also ramped up to $42.0\,\mathrm{G}$, increasing the scattering length to $a\sim800\,a_0$. The following experiments are then performed:

\begin{enumerate}[label=\textbf{\alph*)}]
    \item \textbf{Quench experiment:}
    The long lattices in the staggered configuration are ramped to $V_{\mathrm{l}}\smash{^{(m=0)}}$, corresponding to the $m \approx 0$ resonance ($U \approx 2\Delta$). This depth is calibrated as described in Section~\ref{sec:exp_calibrations}. The short lattice depth is then abruptly reduced to $6\,E_r^{\mathrm{s}}$ in $150\,\text{\textmu s}$, corresponding to a tunnel coupling of $J/h\approx130\,\mathrm{Hz}$. After a variable evolution time $T_{\mathrm{quench}}$, the short lattices are abruptly ramped back to $45\,E_r^{\mathrm{s}}$, further increased to $60\,E_r^{\mathrm{s}}$ over $5\,{\mathrm{ms}}$, and the long lattices simultaneously removed. A doublon-to-singlon conversion step is performed prior to imaging.

    \item \textbf{QSL realisation.}
    
    \textbf{Forward ramp:} The long lattices are initially ramped to $V_{i}\approx0.93V_{\mathrm{l}}\smash{^{(m=0)}}$, corresponding to $m<0$. The short lattice depth is then reduced to $6\,E_r^{\mathrm{s}}$ with a linear ramp over $1.8\,\mathrm{ms}$, followed by a linear ramp on the long lattices with duration $T_\mathrm{forward}$, reaching a final value $V_f\approx1.07V_{\mathrm{l}}\smash{^{(m=0)}}$. Columnar states are prepared by reducing the long lattice depth in one direction by a chemical potential shift $|\mu|$. Finally, the long lattice depth in both directions is jumped to $1.2\,V_l\smash{^{(m=0)}}$, and the atom distribution is then frozen prior to the doublon-to-singlon conversion step.

    \item \textbf{Round-trip.}

    \textbf{Forward ramp:} QSL states are first prepared using a forward ramp as described above.
    
    \textbf{Optional phase imprint step:} The phase imprint between vertical and horizontal dimers is implemented by abruptly jumping the long lattice power to $V_{\mathrm{l}}^x\approx3\,E_r^{\mathrm{l}}$ along the $x$-axis, $V_{\mathrm{l}}^y\approx9\,E_r^{\mathrm{l}}$ along the $y$-axis, and both short lattices to $8\,E_r^{\mathrm{s}}$. After a phase imprinting time $T_{\text{z-pulse}}$, all horizontal lattices are jumped back to their previous values.
    
    \textbf{Reverse ramp:} The reverse sweep is similar to the forward ramp, except that the duration $T_{\mathrm{reverse}}$ may differ. The atom distribution is frozen prior to doublon-to-singlon conversion step.

\end{enumerate}

\subsection{Doublon spilling}\label{sec:doublon_spilling}

Our experiments require simultaneous detection of singlons and doublons in single experimental snapshots. In standard quantum gas microscopy, however, inelastic light-assisted collisions during fluorescence imaging cause doublons to be lost almost immediately, resulting in parity projection of the lattice occupation. We employ a novel vertical spilling method to overcome this limitation and distinguish empty sites from doubly-occupied sites. Similar techniques have been used for deterministic loading of fermions in tightly focused tweezer traps~\cite{serwane_deterministic_2011,jain_programmable_2025}.

We freeze the in-plane dynamics by ramping up the short lattices to $60\,E_r^{\mathrm{s}}$ and tune the scattering length to $a\sim490a_0$ using a Feshbach resonance. The steep-angle vertical lattice (StVL) depth is ramped down from \SI{16}{kHz} to \SI{5.5}{kHz} over \SI{20}{ms}. Gravity leads to a vertical out-of-plane tilt of \SI{3.5}{kHz} per vertical lattice site, allowing atoms to tunnel away at a rate that depends on their energy and thus the barrier height. The strong on-site interaction energy $U$ reduces the effective barrier height for doublons, significantly increasing the doublon-to-singlon conversion rate relative to the singlon loss rate. One can therefore find a range of lattice depths and hold times for which one atom of a doublon tunnels away with high probability while singlons remain trapped, efficiently converting doublons into singlons.

To characterize the fidelity of the gravity-assisted doublon-to-singlon conversion, we apply this technique to initially prepared singlon and doublon charge density waves ($\ket{\cdots 1010\cdots}$ and $\ket{\cdots 2020\cdots}$). Figure~\ref{fig:gauss-law}a of the main text shows the singlon survival probability $p_{1\to1}$ and the doublon-to-singlon conversion probability $p_{2\to1}$ as functions of the StVL depth $V_z$. For $V_{z}/h\lesssim$ \SI{5}{kHz}, both probabilities increase together as the lattice becomes deep enough to hold singlons against gravity. Beyond this threshold, $p_{1\to1}$ saturates while $p_{2\to1}$ exhibits a plateau up to $V_{z}/h\sim$ \SI{6}{kHz} before decreasing at larger depths as both atoms comprising a doublon begin to remain trapped, and parity projection leads to reduced detected filling.

To model this process, we describe the singlon and doublon populations as a function of time using the coupled rate equations
\begin{equation}
\begin{aligned}
    \dot{n}_2(t) &= -\Gamma_2 n_2(t), \\
    \dot{n}_1(t) &= -\Gamma_1 n_1(t) + \Gamma_2 n_2(t).
\end{aligned}
\end{equation}
Here, $n_{1}$ and $n_{2}$ denote the singlon and doublon populations, $\Gamma_{1}$ is the rate at which singlons tunnel away, and $\Gamma_2$ the rate of doublon-to-singlon conversion. We further adopt a phenomenological description of the tunneling rates inspired by Refs.~\cite{gluck_lifetime_1999,zenesini_time-resolved_2009},
\begin{equation}
    \Gamma_r=a_r\exp(-b_rV_z^2),
\end{equation}
where $a_r$ and $b_r$ are free parameters and $r \in \{1, 2\}$ labels singlons and doublons, respectively. These coefficients are extracted by fitting to the experimental data (Fig.~\ref{fig:gauss-law}a of the main text), yielding rates $\Gamma_1 \sim$ \SI{0.1}{Hz} and $\Gamma_2 \sim$ \SI{55}{Hz} at a lattice depth of \SI{5.5}{kHz}. From these parameters, we obtain a singlon survival fidelity of $\mathcal{F}_{1\to1} = 99(1)\%$ and a doublon-to-singlon conversion fidelity of $\mathcal{F}_{2\to1} = 98(1)\%$.

\subsection{Calibration of experimental parameters}
\label{sec:exp_calibrations}

\textbf{Calibration of the staggered superlattice potential $\Delta$ and the on-site interaction $U$:}\\

\begin{figure}[h!]
\includegraphics{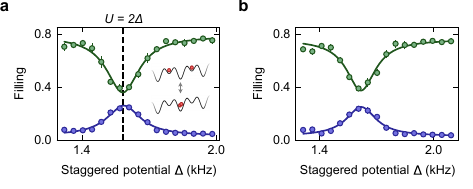}
     \caption{\textbf{Calibration of the kinetically constrained resonance $U \approx 2\Delta$.}
     \textbf{a,} Starting from the monomer initial state, we quench both short lattices and apply an additional staggered potential along the $y$-axis ($\Delta_y-\Delta_x>0$) to restrict the dynamics to the $x$-axis. We measure the vertex filling (green) and doublon-resolved link filling (blue) after \SI{3}{ms} evolution time, observing peak link filling at $U \approx 2\Delta$. From the fit we obtain $U = $\,\SI{1.58(2)}{kHz}.
     \textbf{b,} The same protocol with an additional staggered potential along the $x$-axis ($\Delta_x-\Delta_y>0$) to confine dynamics to the $y$-axis, yielding $U = $\,\SI{1.62(2)}{kHz}.}
     \label{fig:SI_u_calibration}
\end{figure}

The on-site interaction $U$ is calibrated based on the kinetically constrained resonance $U\approx2\Delta$, as shown in Fig.~\ref{fig:SI_u_calibration}. The staggered potential $\Delta$ is independently determined from parametric heating measurements, performed by modulating the lattice amplitude and identifying the resonance frequency at which the atoms are transferred from the first band to the third band. This frequency is then compared with a band-structure calculation to calibrate the absolute depth of the lattice.

We note that, even at $U=2\Delta$, monomers have a small but finite mass due to second-order perturbation theory (see Section~\ref{sec:sm:effective_model}). The resonance condition $m = 0$ calibrated experimentally automatically takes this renormalization into account (Fig.~\ref{fig:SI_u_calibration}).\\

\textbf{Calibration of the tunneling coupling $J$ and the tilt $\delta$:}\\

\begin{figure}[h!]
\includegraphics{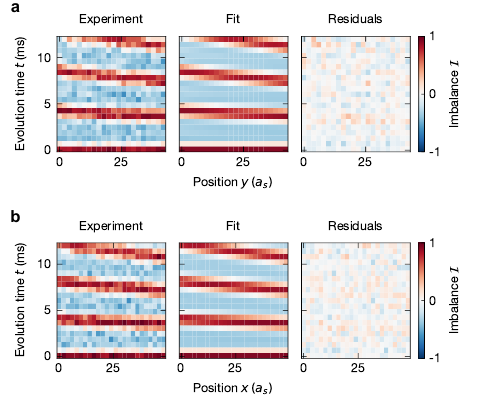}
     \caption{\textbf{Calibration of the tunnelling coupling $J$ and the tilt $\delta$ using Bloch oscillations.} \textbf{a,} Time evolution of the imbalance $\mathcal{I}$ starting from a charge density wave initial state along the $x$-axis. Each time data point represents an average over three experimental realisations, further averaging over the perpendicular direction. The middle panel shows the fitted model obtained from Eq.~\eqref{Eq:Imbalance_Bloch_osc}, and right panel the fit residuals. \textbf{b,} Same measurement and analysis performed along the $y$-axis.}
     \label{fig:SI_bloch_oscillations}
\end{figure}

The overall on-site potential experienced by the atoms, in addition to the staggered superlattice potential, consists of a diagonal tilt due to a magnetic field gradient and a smaller contribution from harmonic confinement of the vertical lattice (and to a smaller extent horizontal superlattices). Along individual lattice axes this potential is described by
\begin{equation}
    V(x) = \delta_0\frac{x}{a_s}+ \frac{1}{2}m\omega^2(x-x_0)^2,
\end{equation}
where $x$ is the position along the axis, $\delta_0$ is the tilt due to the magnetic field gradient, $m$ is the atomic mass of caesium, $\omega$ is the in-plane harmonic trap frequency, and $x_0$ is the trap centre. This gives a spatially varying tilt $\delta(x)/a_s = \delta_0/a_s+m\omega^2(x-x_0)$. To calibrate $J$ and $\delta(x)$ along each lattice axis, we study Bloch oscillations starting from a charge density wave initial state $\ket{\cdots101010\cdots}$ (Fig.~\ref{fig:SI_bloch_oscillations}), in the absence of the staggered potential ($\Delta = 0$). The on-site interaction $U$ is set to be $U \gg J$ for this experiment, such that we can treat the bosons as free fermions. The time evolution of the imbalance as a function of spatial coordinate $x$ along the chain is given by~\cite{scherg_observing_2021}
\begin{equation}
    \mathcal{I}(x;t) = A\mathcal{J}_0\left(\frac{8J}{\delta(x)}\sin\left(\frac{\delta(x) t}{2\hbar} \right)\right)e^{-t/\tau},
    \label{Eq:Imbalance_Bloch_osc}
\end{equation}
where $\mathcal{J}_0$ denotes the zeroth-order Bessel function of the first kind, $A$ is a constant pre-factor that accounts for imperfect initial state filling, and $\tau$ is an additional damping factor that accounts for dephasing due to on-site potential disorder. The extracted fit parameters for both lattice axes are shown in Table~\ref{tab:fit_results_bloch_osc}.

\begin{table}[h]
\centering
\begin{tabular}{c | c c c c c}
 &\, $J/h$ (Hz) & \,$\delta_{\min}/h$ (Hz)& \,$\delta_{\max}/h$ (Hz) \,& $\omega/2\pi$ (Hz) \\
\hline
$x$-axis \,& 132(1) & 234(4) & 267(4) & 20.1(6) \\
$y$-axis \,& 129(1) & 246(3) & 279(4) & 18.8(5) \\
\end{tabular}
\caption{\textbf{Parameters extracted from the Bloch oscillation fits.} Parameters extracted from the fits shown in Fig.~\ref{fig:SI_bloch_oscillations}. The fit amplitude is $A=0.94(2)$ along both axes, and the exponential decay constants are $\tau_x=80(20)$ ms and $\tau_y=90(20)$ ms.}
\label{tab:fit_results_bloch_osc}
\end{table}


\section{Theory}\label{sec:suppmat:theory}

Gauge theories are convenient theoretical descriptions of correlated quantum states with intrinsic topological order. The cost of working with such theories is the introduction of redundancies in the form of an extended Hilbert space. The presence of a gauge symmetry is necessary to ensure that gauge-variant (“unphysical”) states, i.e., those violating Gauss’s law, cannot be reached. The most straightforward way to eliminate these states is to enforce Gauss’s law with a large energy penalty. However, this typically requires implementing high-body operators and is a key engineering challenge for most synthetic quantum platforms.

To circumvent these problems, we take a different route to realize gauge symmetries: rather than enforcing Gauss’s law energetically, we engineer the dynamics such that, for carefully chosen initial states, time evolution respects Gauss’s law. In the language of Hilbert-space fragmentation~\cite{sala_ergodicity_2020,khemani_localization_2020,moudgalya_quantum_2022}, local dynamical constraints partition the Hilbert space into disconnected sectors. Once prepared, the system remains in a gauge-symmetric sector of the full Hilbert space. In this sense, the fragmentation of the Hilbert space into dynamically disconnected pieces yields a dynamically preserved Gauss’s law, even though gauge-variant states exist at comparable energies.

\subsection{Effective model}\label{sec:sm:effective_model}

\begin{figure*}
\includegraphics{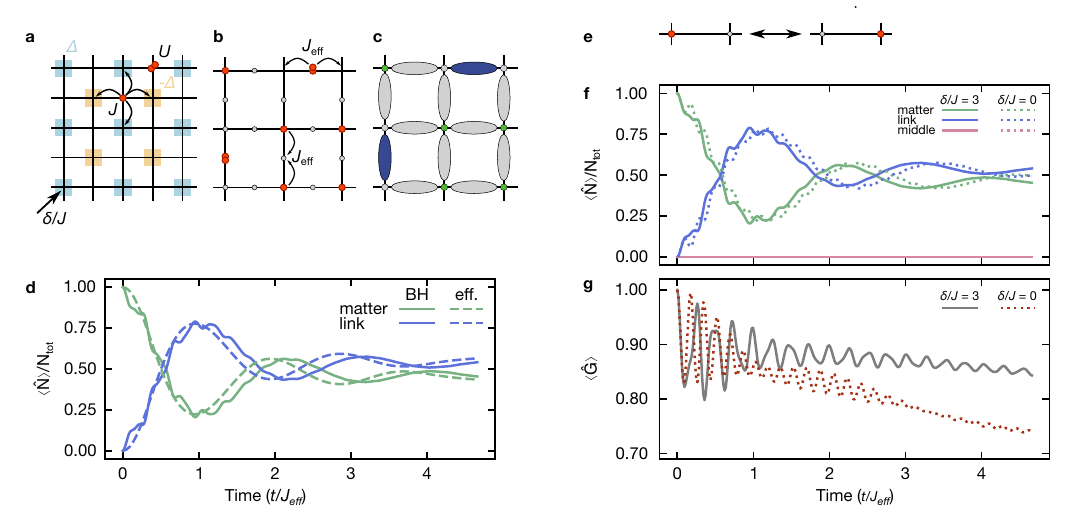}
     \caption{\textbf{Mapping the Hubbard model to a $U(1)$ gauge theory.} \textbf{a} Schematic of
the tilted, two dimensional Bose-Hubbard model with a tilted potential of strength $\delta$ (indicated by the large arrow), tunnelling $J$ and Hubbard interaction $U$ as well as a staggered on-site potential $\Delta$. In the limit $\Delta = U/2 > \delta$ and $U \gg J$ the effective Hamiltonian is fragmented and realizes a gauge theory in certain sectors. \textbf{b} The sector connected to all matter sites (vertices) occupied undergoes constrained dynamics on a Lieb-lattice, which conserve a Gauss law: two neighbouring singlons can form a doublon via an effective coupling strength $J_{\text{eff}}$. In addition, the effective Hamiltonian also has an on-site potential of strength $\delta_m$ on matter sites and $\delta_l$ on links. Red filled circles correspond to a boson, whereas a grey dot, marks an empty site. \textbf{c} The effective model can be mapped to a 2D monomer-dimer model, with a natural $U(1)$ gauge theory description. Singlons are green monomers and doublons blue dimers.
\textbf{d} Time evolution under the full Bose-Hubbard model (solid lines) and the effective model (dashed lines) shows good agreement following a quench from the all-matter initial state with open boundary conditions. MPS simulation parameters: $\chi = 256, dt = 0.001, J=1, U=20, \Delta = 10, \delta = 3$.
\textbf{e} In the absence of the linear potential, the effective Hamiltonian additionally contains the hopping process shown here: singlons can hop from matter site to neighbouring matter sites. This process violates Gauss's law. The applied tilt suppresses this process, thereby stabilizing the gauge-invariant dynamics.
\textbf{f} In the absence of the linear potential, the effective model contains additional hopping processes that violate Gauss’s law. Despite this, the mean occupations of matter and link sites oscillate similarly in both cases and the middle site remains effectively blocked. \textbf{g} The additional processes give rise to an enhanced deviation of the Gauss-law expectation value from one in the no-tilt quench (dotted line) than in the presence of a tilt (solid line), where those processes are absent. Parameters: $\chi = 256, dt = 0.001, J = 1, U= 20,\Delta = 10$.}
     \label{fig:SI_setup_quench}
\end{figure*}

We engineer such a fragmented Hamiltonian, starting from a strongly coupled Bose-Hubbard Hamiltonian on a square lattice,
\begin{equation}
    \begin{aligned}
        \hat H &= - J\sum_{\langle i,j\rangle} \left( \hat a_{i}^{\dagger} \hat a_{j}^{\phantom{\ast}} + \mathrm{h.c.} \right) + \frac{U}{2} \sum_{i} \hat n_{i} (\hat n_{i}-1)\\ & + \sum_{i = (i_x, i_y)} \left[ \frac{(-1)^{i_x} + (-1)^{i_y}}{2} \Delta + (i_x+i_y)\delta \right]\hat n_{i},
    \end{aligned}
    \label{eq:bhm_hamiltonian}
\end{equation}
where $\hat a_{i}^{(\dagger)}$ annihilates (creates) a boson on site $i$, $J$ denotes the tunnelling amplitude, and $U$ is the on-site interaction strength. We further apply a site-dependent on-site potential, consisting of a linear tilt along the lattice diagonal with strength $\delta$ and a staggered potential $\Delta$, see Figure~\ref{fig:SI_setup_quench}a for a schematic illustration. 

In the limit of $U\gg J$, $\Delta = \frac{U}{2}>\delta$, second order perturbation theory yields correlated hopping processes that fragment the Hilbert space of the effective Hamiltonian into disconnected sectors. Within this fragmented Hamiltonian, the `all matter' state (see Fig.~\ref{fig:gauss-law}c) lies in a sector that preserves a Gauss law. In this limit, the yellow sites, illustrated in Figure~\ref{fig:SI_setup_quench}a (in the following referred to as middle sites) are energetically blocked and remain empty under dynamics. This gives rise to the Lieb-lattice structure illustrated in Figure~\ref{fig:SI_setup_quench}b. The sites marked in blue in Figure~\ref{fig:SI_setup_quench}a correspond to the matter sites and are the vertices of the Lieb-lattice in Figure~\ref{fig:SI_setup_quench}b. The effective Hamiltonian governing the dynamics in the all matter state sector is determined by the correlated hopping processes shown in Figure~\ref{fig:SI_setup_quench}b. Specifically, singlons on neighboring matter sites can form doublons on links with hopping strength $J_{\text{eff}}$. Additionally, there is an on-site potential $\delta_m$ on matter sites and $\delta_l$ on link sites. The effective Hamiltonian on periodic boundary conditions is then given by

\begin{equation}
    \begin{aligned}
     \hat{H}_{\text{eff}} &
     = J_{\text{eff}}\sum_{l}
  \left( |{\includegraphics[height=0.2cm]{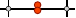}}\rangle\langle{\includegraphics[height=0.2cm]{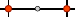}}|+h.c \right) \\
 &+ \delta_m \sum_v |{\includegraphics[height=0.2cm]{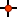}}\rangle\langle{\includegraphics[height=0.2cm]{figures/SI/fig_SI_onsite_vertex.pdf}}| + \delta_l \sum_{l} |{\includegraphics[height=0.2cm]{figures/SI/fig_SI_hop2_hor.pdf}}\rangle\langle{\includegraphics[height=0.2cm]{figures/SI/fig_SI_hop2_hor.pdf}}|
 \end{aligned}
 \label{eq:eff_hamiltonian}
\end{equation}
where $v$ refers to all vertices (matter) sites, and $l$ labels all horizontal as well as vertical links. The effective couplings can be expressed in terms of the parameters of the original Bose-Hubbard model
\begin{align}
    &J_{\text{eff}}=\sqrt{2} \left(\frac{J^2}{\frac{U}{2}-\delta}+ \frac{J^2}{\frac{U}{2}+\delta} \right) \\
    &\delta_m = \frac{2J^2}{\frac{U}{2}-\delta}+\frac{2J^2}{\frac{U}{2}+\delta} \\
    &\delta_l = \frac{2J^2}{\frac{U}{2}-\delta}+\frac{2J^2}{\frac{U}{2}+\delta} + \frac{2J^2}{\frac{3U}{2}-\delta}+\frac{2J^2}{\frac{3U}{2}+\delta} .
\end{align}

In the language of the 2D monomer-dimer model, singlons on matter sites correspond to monomers and doublons on links to dimers. We give an illustrative example of the state in Figure~\ref{fig:SI_setup_quench}b and c. In this language, the resulting effective Hamiltonian under periodic boundary conditions in the sector of the all matter state is given by 
\begin{equation}
    \begin{aligned}
\hat{H}_{\text{eff}} &= J_{\text{eff}} \sum_{\text{links}} \left( \horDBket \horMBbra + \text{h.c.} \right) \\
&+ (\delta_m-\delta_l/2)\sum_{\text{vertices}}\horMket \horMbra ,
    \end{aligned}
    \label{eq:sm:monomer-dimer}
\end{equation}
where $\horMket$ denotes a monomer, $\horDket $ a dimer and $\horMEket $ and $\horDEket $ are empty sites. 
The local Gauss's law can be formulated as 

\begin{equation}
\hat{G}_V \equiv \horMket \horMbra+ \sum_{\mathrm{links\,at\,}V} \horDket \horDbra =1.
\end{equation}

The equivalence to a $U(1)$ lattice gauge theory can be made explicit by rewriting the Gauss law in a more conventional form in the charge-free sector
\begin{equation}
    \mathrm{div}\,\hat{\mathbf{E}}(\mathbf{r}) = 0.
     \label{eq:divE}
\end{equation}
To this end, we introduce the lattice divergence as $\mathrm{div}\,\hat{\mathbf{E}}(\mathbf{r}) = \Delta_x \hat{E}_x(\mathbf{r})+\Delta_y\hat{E}_y(\mathbf{r}) $, with
\begin{equation}
\Delta_x\hat{E}_x(\mathbf{r}) = 
\hat{E}_x(\mathbf{r}+\frac{1}{2}e_x)-\hat{E}_x(\mathbf{r}-\frac{1}{2}e_x)
\end{equation}
defined as a finite difference with matter-sites are labelled by position $\mathbf{r}$. In the gauge theory subspace, we can express the electric fields in terms of the Bose-Hubbard boson operators $\hat{a},\hat{a}^\dagger$ as follows: 
\begin{equation}
\begin{aligned}
    \hat{E}_{x}(\mathbf{r}) &= (-1)^{r_x+r_y}\left[2\; \hat{a}^\dagger \hat{a}(\mathbf{r}+\frac{1}{2}\mathbf{e}_x)-1\right] \\
    \hat{E}_{y}(\mathbf{r}) &= (-1)^{r_x+r_y}\left[2\; \hat{a}^\dagger \hat{a}(\mathbf{r}+\frac{1}{2}\mathbf{e}_y)-1\right].
\end{aligned}
\end{equation}
Electric field operators are associated to a site $\mathbf{r}$, but act only the links of the Lieb lattice. The boson occupations on the links take values in $\hat{a}^\dagger \hat{a} \in \lbrace 0,2 \rbrace$ if we are constrained to the physical subspace.

\subsection{Equivalent descriptions of the effective model}
The effective model in Eq.~\ref{eq:sm:monomer-dimer} implements a $U(1)$ lattice gauge theory. The explicit gauge redundancy present in our model and the experiments, must be fixed by imposing Gauss law $\hat{G}_V=1, \forall V$ on physical states. It is also possible to construct an equivalent Hamiltonian without gauge redundancies, which acts identical to Eq.~\ref{eq:sm:monomer-dimer} on physical states. By resolving the Gauss-law to track matter excitations using the electric field divergence $\mathrm{div}\,\hat{\mathbf{E}}\neq0$, we obtain a model which acts only on effective spin-1/2 degrees of freedom on the links. To make the notation simpler, we choose to represent
\begin{equation}
    \horDket  \text{ as } \horUPket \qquad \text{and} \qquad \horDEket \text{ as } \horDOWNket. 
\end{equation}
The resulting model takes a `PXP' form, with
\begin{equation}
\tilde{H}_{\mathrm{eff.}} = J_{\text{eff}}\sum_{l\in\text{links}} \hat{P}_{} X_l \hat{P} - (\frac{1}{2}\delta_l-\delta_m) \sum_{l\in\text{links}} Z_l,
\end{equation}
up to a constant energy shift. The projector $\hat{P}$ ensures that any links sharing a site with $l$ are empty. While the `PXP' Hamiltonian presents simpler, the gauge-theory description is crucial to implement this Hamiltonian in our quantum simulator. The gauge description avoids implementing projection operators involving multiple sites which is prohibitively complex to explicitly realize experimentally.

\subsection{Quench}

We investigate the dynamics of the system starting from an initial state with singlons on all matter sites and compare the time evolution of the state under both the full Bose-Hubbard Hamiltonian and the effective Hamiltonian. In Figure~\ref{fig:SI_setup_quench}d we show average number of bosons on matter sites (green) and the average link occupation (blue) for parameters $J = 1, U = 20, \delta = 3$. Dashed lines correspond to time evolution under the effective Hamiltonian, while solid lines show results obtained from the full Bose-Hubbard model, simulated using the Python library TeNPy \cite{hauschild_tensor_2024}.
We have verified convergence in bond dimension $\chi$ and time step size $dt$ by comparing simulations with $\chi = 512, dt = 0.0001$ versus $\chi = 256, dt = 0.001$, finding excellent agreement between the two. Our results show a good agreement between the two Hamiltonians, which demonstrates the effectiveness of the strong coupling approach in describing dynamics of the system.

The Gauss law is dynamically conserved since violating processes, such as the ones illustrated in Figure~\ref{fig:SI_setup_quench}e, are suppressed by appropriate choice of parameters of the Bose-Hubbard Hamiltonian, including the presence of a linear tilt.
To illustrate this, we additionally compare dynamics under the Bose-Hubbard Hamiltonian with and without the tilt. 
Remarkably, Figure~\ref{fig:SI_setup_quench}f shows that although occupations of matter and link sites remain similar in both cases, the validity of Gauss's law deteriorates over time in absence of the tilt, see Figure~\ref{fig:SI_setup_quench}g. Starting from an initial state in which all matter sites are singly occupied, the process shown in Figure~\ref{fig:SI_setup_quench}e is not immediately possible. Instead, the system must first undergo dynamics to generate empty matter sites before such a process can occur. We attribute the delayed onset of Gauss-law violations in the un-tilted case to this fact. This demonstrates that fragmentation plays a crucial role in suppressing gauge-violating processes and stabilizing the effective gauge theory.
We remark that in both cases the middle site of the plaquette stays unoccupied, justifying the Lieb-lattice structure of the effective Hamiltonian, see Figure~\ref{fig:SI_setup_quench}f.

\subsection{Energy spectrum}

\begin{figure*}
\includegraphics{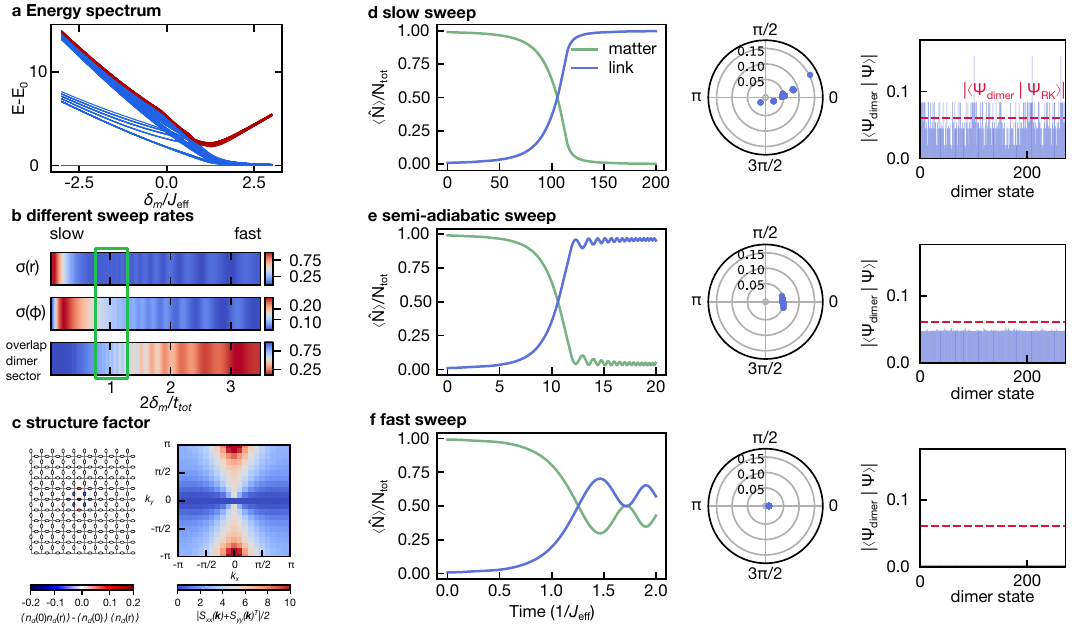}
     \caption{\textbf{a} \textbf{Spectrum of effective model.}  Lowest eigenstates for periodic boundary conditions, Parameters: $J_{\text{eff}} = 1, \delta_m \in [-3,3]$, system size $4\times4$ matter sites. \textbf{b} Analysing the final state after the sweep for different sweep rates, reveals three regimes. The sought after intermediate regime realizes small standard deviation in amplitudes of dimer states, as well as small circular standard deviation of their phases while having non-vanishing overlap with the dimer sector. \textbf{c} Classical simulation  of the structure factor and its Fourier transform in the pure dimer sector. The Fourier transform, reveals pinch points. Colorscale is saturated for $(k_x,k_y)=(0,\pi)$. left: $N_{updates} = 100000, N_{av} = 10000, L_x,L_y = 20$, right: $N_{updates} = 500000, N_{av} = 3000, L_x,L_y = 40$. \textbf{d-f Sweep in the effective model. } Starting from the groundstate of the effective model with $\delta_l = 0$ and $\delta_m = -10$, the matter onsite potential $\delta_m$ is tuned linearly over time with a rate $r = \frac{2\delta_m}{t_{tot}}$ to $\delta_m = 10$. Matter and link occupations are shown over time for different sweep rates, as well as the overlap with the dimer manifold in the polar plots. For a slow ramp, shown in \textbf{d} with $t_{tot} = 10 (1/J_{\text{eff}})$, the final state has particles dominantly on links, but dimer states are not an equal weighted coherent superposition. \textbf{e} In the semi-adiabtic case $t_{tot} = 1 (1/J_{\text{eff}})$, the final state has most particles on links, and dimer states have approximately equal weight and phase in the final state. For fast sweeps \textbf{f} shows that the initial state will leave the low energy manifold over time and links as well as matter sites will be occupied. Furthermore there is low weight on dimer states. Parameters: $J_{\text{eff}}= 1, \delta_m \in [-10,10], dt = 0.025$.}
     \label{fig:SI_sweep}
\end{figure*}

Figure~\ref{fig:SI_sweep}a shows the spectrum of the effective model with periodic boundary conditions for $J_{\text{eff}} = 1,\ \delta_l = 0$ and varying onsite potential strength on matter sites $\delta_m$. For $\delta_m \ll 0 $, the ground state is approximately a unique product state of all-matter sites singly occupied, which can be understood as a \textit{Higgs phase} in the gauge theory. The first excited states are separated by a gap of roughly $2\delta_m$ and correspond to dimer excitations. In the opposite limit, $\delta_m \gg 0$, many low-energy states appear, which are well approximated by the set of all dimer configurations fulfilling Gauss's law as in Eq.~\eqref{eq:divE}. This additional constraint of vanishing divergence further constrains the dynamics in the dimer manifold, which is then additionally suppressed by a factor $\mathcal{O}(J_{\text{eff}}^4/\delta_m^3)$. The small energy gaps in this regime make the adiabatic preparation of the ground state for $\delta_m \gg 0$ challenging. To further complicate the analysis, the generic fate of the ground state at $\delta_m /J_{\text{eff}}\gg1 $ will be that the gauge theory forms a \textit{confined phase} by spontaneous symmetry breaking in the dimer manifold~\cite{moessner_quantum_2008, fradkin_field_2013}. Deconfined spin liquid phases still exist, and in the case of the monomer-dimer model, take the form of coherent equal-weight superpositions of all dimer coverings of the square lattice. However, in thermal equilibrium, such states are only stable as isolated critical points in the phase diagram~\cite{polyakov_1977,rokhsar_superconductivity_1988}.

\subsection{Sweep}
Despite the difficulty in the adiabatic preparation of deconfined states, such spin-liquid-like states may still be prepared by starting in the Higgs phase for $\delta_m/J_{\text{eff}} \ll -1$ and then sweeping $\delta_m$ from negative to positive values. Concretely, we initialize the system in the ground state of the effective Hamiltonian \eqref{eq:eff_hamiltonian} with $\delta_m/J_{\text{eff}}= -10$ and $\delta_l/J_{\text{eff}} = 0$, and then ramp the onsite potential $\delta_m$ linearly in time up to $\delta_m/J_{\text{eff}} = 10$. The desired final state after the sweep should approximate the Rokhsar-Kivelson (RK) state as closely as possible. In particular, it should have equal amplitudes and coherent phases among all dimer configurations while minimizing the amount of matter excitations. To quantify this, we analyze three observables: First, the relative standard deviation of the normalized amplitudes of dimer-states. Second, the circular standard deviation of their phases, and third, the total overlap with the dimer subspace. The circular spread is quantified by the circular standard deviation 
\begin{align}
    \sigma_{circ} &= \sqrt{-2\ln R_w}; \\
    R_w &= |\sum_l r_l e^{i\Phi_l}|,
\end{align}
where $r_l$ are the amplitudes of dimer states in the final states, normalized to one, and $\Phi_l$ are their respective phases. As shown in Fig.~\ref{fig:SI_sweep}b different sweep rates $r = \frac{2\delta_m}{t_{tot}}$, can be sorted into too slow, too fast, and an intermediate regime in which both amplitude and phase fluctuations are suppressed, and the overlap with the dimer sector remains finite.

For sufficiently slow sweeps, even compared to the finite-size gaps in the system, we are approaching an adiabatic regime. By analysing the occupations of matter and links over time, we observe an absence of matter excitations at the end of the sweep when $\delta_m \gg J_{\text{eff}}$, see Fig.~\ref{fig:SI_sweep}d. However, as shown in the polar plot of Figure~\ref{fig:SI_sweep}d, the final state has different weights on individual dimer states, which indicates the tendency of the system to form a confined phase. This is further illustrated in the right plot, which shows the absolute value of the overlap of all possible dimer states with the RK wave function. An ideal RK state would correspond to the red dashed line, representing equal weight across all dimer configurations. Ramping slowly, we therefore find that the weight of dimer states is not equally distributed.

In the opposite regime, the sweep is too fast for the system to follow the low-energy manifold, and higher energy states are also populated. Consequently, we find a significant fraction of residual matter excitations; see Figure~\ref{fig:SI_sweep}f. Furthermore, there is little weight on dimer states in the final state after the sweep.

The desired regime is the semi-adiabatic one: the sweep is as adiabatic as possible with respect to creating matter excitations (shown in red in Figure~\ref{fig:SI_sweep}a), but diabatic with respect to the small splittings within the dimer manifold. In this case, the system remains mostly in the low-energy dimer subspace, while the final state is given by an approximately equal weighted, coherent superposition of dimer states, indicative of a deconfined RK spin liquid; see Figure~\ref{fig:SI_sweep}e. We remark that in the thermodynamic limit, the many-body gap closes, which means that a truly adiabatic regime is impossible. Thus, we will always create a finite density of matter excitations via Kibble-Zurek physics. The residual density of matter excitations $\rho_m$ then sets an upper bound on the sizes of spin-liquid regions, which in turn depends on the nature of the critical point.

\begin{figure}[h!]
\includegraphics{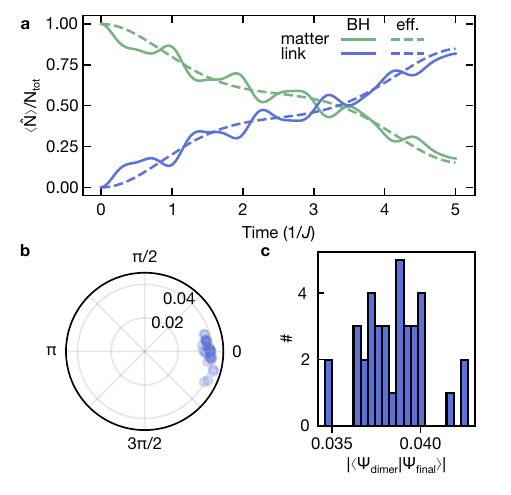}
     \caption{\textbf{Sweep in the Bose Hubbard model.} Starting from the all-matter state, an additional onsite potential $\nu$ on matter sites is ramped linearly in time from negative to positive values. Dashed lines show the corresponding sweep for the effective model with open boundary conditions. During the sweep, particles are transferred from matter to link sites. \textbf{b} The final state exhibits a finite overlap with the pure dimer sector. The amplitudes and phases of the dimer components are shown in the polar plot, while a histogram of the amplitudes is shown in \textbf{c}. MPS simulation parameters: $L_x = L_y = 4, \chi = 256, dt = 0.005, J=1, U=20, \Delta = 10, \delta = 3, \nu \in [-1,1], t_{sweep} = 5$.}
     \label{fig:SI_sweep_ef_vs_mps}
\end{figure}
We checked that similar conclusions also hold in the microscopic Bose-Hubbard model by performing MPS based simulations on system sizes of $4\times 4$ matter sites. Figure~\ref{fig:SI_sweep_ef_vs_mps}a shows MPS data for parameters that are closest to experimental data. At the end of the sweep, about $83\%$ of matter has been converted to dimers. Figure~\ref{fig:SI_sweep_ef_vs_mps}b shows angles and amplitudes of pure dimer states in the final wavefunction in a polar plot and the distribution of amplitudes in a histogram in Figure~\ref{fig:SI_sweep_ef_vs_mps}c. We compare the evolution of particle number over time  with the effective model under open boundary conditions and find good agreement, see Figure~\ref{fig:SI_sweep_ef_vs_mps}a. 

\subsection{Signatures of the final state}
At the end of the sweep, the system reaches a configuration with very few matter excitations, thus realizing an approximately pure gauge theory, in which
\begin{equation}
\hat{G}_V \approx \sum_{\mathrm{links\,at\,}V} \horDket \horDbra =1.
\end{equation}
or equivalently 
\begin{equation}
    \mathrm{div}\,\mathbf{E}(\mathbf{r}) \approx 0.
\end{equation}
The absence of matter strongly affects electric field correlations. This shows up in the static electric field structure factor
\begin{equation}
    S_{\alpha\beta}(\mathbf{k}) = \langle \hat{E}_\alpha(\mathbf{k}) \hat{E}_\beta(-\mathbf{k})\rangle,
\end{equation}
which develops a characteristic pinch-point-like behaviour due to the transversal nature of the electric fields. Specifically, translational invariance and rotational symmetry enforce the general form
\begin{equation}
    S_{\alpha\beta}(\mathbf{k}) \sim \left(\delta_{\alpha\beta}-\frac{k_\alpha k_\beta}{\mathbf{k^2}}\right),
\end{equation}
for low momenta, which is clearly visible in the numerics (Fig.~\ref{fig:SI_sweep}c), and the experiment (Fig.~4d in the main text).

\subsection{Detecting coherence}
\begin{figure}
\includegraphics{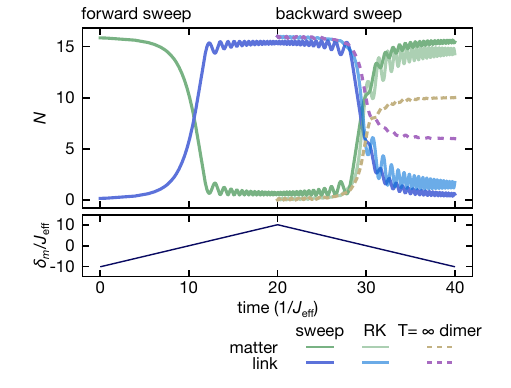}
     \caption{\textbf{Round-trip protocol probing coherence.} Starting from the groundstate of the effective model with $\delta_l = 0$ and $\delta_m = -10$, the matter onsite potential $\delta_m$ is tuned linearly over time with a rate $r = \frac{2\delta_m}{t_{tot}}$ converting matter into dimers. During the backward sweep $\delta_m$ is linearly decreased back to its original value, reconverting dimers into matter. The RK wavefunction shows the same behaviour undergoing the backward sweep. In contrast, starting from an infinite temperature state in the dimer manifold converts back into a high density of dimer excitations. Parameters: $\delta_m / J_{\text{eff}} \in [-10 ,10]$, $J_{\text{eff}} = 1$, $dt = 0.025$.}
     \label{fig:SI_backward_sweep}
\end{figure}

To probe coherence in the wavefunction after the forward sweep, we implement a round-trip protocol consisting of the forward sweep followed by a backward sweep in which the ramp of the matter potential $\delta_m$ is reversed, see Fig.~\ref{fig:SI_backward_sweep}. For the full round-trip evolution, we find that matter sites are predominantly reoccupied at the end of the protocol. This behaviour aligns with the backward evolution of a RK wavefunction.
This evolution under the backward sweep is in stark contrast to the backward evolution of an incoherent superposition of dimers. As shown in  Fig.~\ref{fig:SI_backward_sweep}, an infinite temperature state in the dimer manifold converts back into a high density of dimer excitations. We have checked this by evolving all individual dimer-states backward and averaging the outcome.

\section{Data analysis}

\subsection{Extended pattern analysis in the quench experiment}

\begin{figure}[h!]
\includegraphics{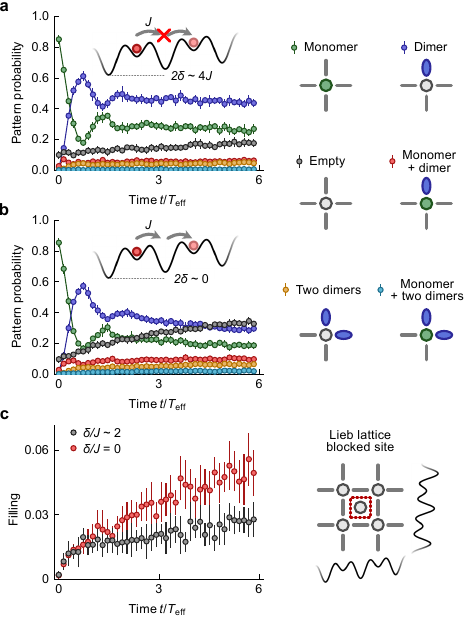}
     \caption{\textbf{Time evolution of vertex pattern probabilities and blocked-site occupation during the quench experiment.}
     \textbf{a,} Detection probabilities of the six most common vertex patterns as a function of time after the quench from the all-monomer state, with diagonal tilt $\delta/J \sim 2$. The first two patterns satisfy Gauss's law; the remaining four are violations.
     \textbf{b,} Same as panel~(a), without the tilt.
     \textbf{c,} Time evolution of the blocked-site filling with (grey) and without (red) tilt. Each data point is averaged over 15 experimental realisations; error bars denote one standard deviation.
     }
     \label{fig:all_patterns_quench}
\end{figure}

Figs.~\ref{fig:all_patterns_quench}a,b show the detection probabilities of the six most common vertex patterns as a function of time after the quench, with and without the tilt. The defects in the initial state are predominantly missing monomers (empty-vertex pattern). The early-time dynamics of the matter and gauge degrees of freedom are strikingly similar in the presence and absence of the tilt; however, the fraction of the two Gauss's-law-allowed patterns decreases significantly faster in the absence of the tilt, as monomers hop directly between vertices. This Gauss's-law-violating process triggers an avalanche of secondary processes. When a vertex is vacated by dimer formation and a monomer from a neighbouring vertex hops directly onto it, a monomer-dimer pattern and an empty-vertex pattern appear. The proximity of two such patterns can then trigger a double-dimer formation. This mechanism is visible even in the tilted case, indicating that residual direct monomer hopping persists, and is more pronounced in the absence of the tilt.

The occupation of sites outside the effective Lieb lattice (blocked sites; Fig.~\ref{fig:all_patterns_quench}c) remains below $3\%$ in the tilted case but increases significantly faster without the tilt. One process that populates the blocked sites follows from the monomer-dimer pattern: the doublon on the link can break, placing a doublon on the vertex while the remaining atom hops onto the blocked site.

\subsection{Hilbert space fragment dimension extraction}

Following Ref.~\cite{karch_probing_2025}, the effective Hilbert space dimension explored during the dynamics can be quantified using the subsystem return probability $\mathcal{L}_A$, which serves as a proxy for the global return probability while offering practical experimental advantages such as robustness to atom loss. In brief, it measures the probability of finding an initial-state patch of area $A$ in the time-evolved snapshots. For a subsystem of $A$ vertices, starting from a translation-invariant initial state, it is defined as
\begin{equation}
    \mathcal{L}_A = \left\langle\prod_{V\in A} \hat{P}_{V}\right\rangle,
\end{equation}
where $\hat{P}_V$ projects onto the initial monomer state at vertex $V$: it equals one if $V$ is occupied by a monomer with no dimers attached and zero otherwise.

In the long-time regime, the time-averaged subsystem return probability $\overline{\mathcal{L}_A}$ quantifies the effective Hilbert space dimension $\dim\mathcal{H}_{\mathrm{eff}}$ via
\begin{equation}(\dim\mathcal{H}_{\mathrm{eff}})^{1/\mathcal{A}}\approx\exp\left(\diff{(-\ln\overline{\mathcal{L}_A})}{A}\right),
\end{equation}
where $\mathcal{A}\to \infty$ is the total number of vertices in the thermodynamic limit. In practice, the subsystem area used to extract the slope must be large enough to capture all relevant correlations.

\begin{figure}[h!]
\includegraphics{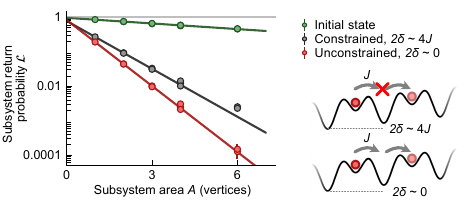}
     \caption{\textbf{Subsystem return probability in the quench experiment.} Subsystem return probability $\mathcal{L}_A$ as a function of the subsystem area $A$ for the initial state (green), and in the quench experiment in the constrained (grey) and unconstrained (red) cases. Each data point represents an average over rectangular subsystems of the same shape, over 86 experimental realisations for the initial state, and over 13 evolution times with $4\lesssim t/T_{\mathrm{eff}}\lesssim6$ and approximately eight realisations per data point for the time-evolved states. Quasi-1D rectangles whose sides differ by more than 3 vertices are not considered in the analysis. Error bars are estimated using standard error of proportion; where not visible, are smaller than the markers.}
     \label{fig:fig_sm_srp_quench}
\end{figure}

In our experiment, evolution times are limited to $6\hbar/J_{\mathrm{eff}}$, but since the local dynamics equilibrate within $\sim 2\hbar/J_{\mathrm{eff}}$, we can extract an approximate value of $\dim\mathcal{H}_{\mathrm{eff}}$. In Fig.~\ref{fig:fig_sm_srp_quench} we plot $\overline{\mathcal{L}_A}$ (averaged over the time window $t>4\hbar/J_{\mathrm{eff}}$) as a function of subsystem area $A$ for the constrained ($\delta/J\sim 2$) and unconstrained ($\delta/J = 0$) gauge theory, from which we extract the values of $\dim\mathcal{H}_{\mathrm{eff}}$ quoted in the main text.

We note that this estimate is approximate for two reasons. First, accessible evolution times are insufficient for full thermalisation. Second, Gauss's law enforces non-zero correlations even in a fully thermalised state at infinite temperature, so the subsystem return probability does not reach the exponential regime until the subsystem is large enough to capture all relevant correlations. With the number of snapshots available, we cannot reliably access this regime, introducing a systematic error in our estimate.

\subsection{Postselection of QSL snapshots on the dimer imbalance}

The experimental snapshots of the QSL state after the forward ramp (Fig.~\ref{fig:rk_experiment}) are postselected on dimer imbalance between $-20\%$ and $+20\%$, retaining 431 out of 877 snapshots. The dimer imbalance is defined as
\begin{equation}
    \frac{N_H-N_V}{N_H+N_V},
\end{equation}
where $N_{H}$ and $N_V$ are the numbers of horizontal and vertical dimers, respectively. A non-zero imbalance develops when the two lattice axes are not perfectly symmetric, for instance, due to pointing drifts of the lattice laser beams caused by temperature fluctuations on the optical table.

\subsection{Pinch points}

\begin{figure}[h!]
\includegraphics{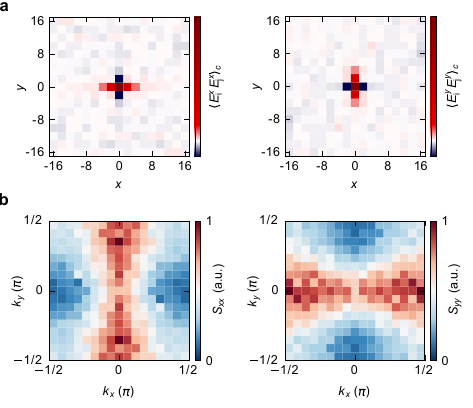}
     \caption{\textbf{Electric field correlations and pinch points.}
     \textbf{a,} Connected electric field correlations $\langle \vec{E}_\alpha(\mathbf{r_i}) \vec{E}_\alpha(\mathbf{r_j})\rangle_c$ as a function of separation $(x,y) = \mathbf{r}_i - \mathbf{r}_j$, evaluated assuming spatial inversion symmetry.
     \textbf{b,} Electric field structure factors $S_{xx}$ and $S_{yy}$, computed as the 2D Fourier transform of the correlations in panel~(a), exhibiting characteristic pinch points at $\mathbf{k} = \mathbf{0}$. Results are averaged over 431 experimental realisations.
     }
     \label{fig:fig_sm_pinchpoints}
\end{figure}

The electric field structure factor $S_{\alpha\beta}$ for $\alpha = \beta$ (where $\alpha = x, y$) is given by
\begin{equation}
    S_{\alpha\alpha}(\mathbf{k}) = \langle \vec{\tilde E}_\alpha(\mathbf{k}) \vec{\tilde E}_\alpha(-\mathbf{k})\rangle_c = \langle | \vec{\tilde E}_\alpha(\mathbf{k}) |^2 \rangle_c,
\end{equation}
where $\vec{\tilde E}_\alpha(\mathbf{k}) = \sum_i \vec{E}_\alpha(\mathbf{r_i}) e^{-i\mathbf{k}\cdot\mathbf{r_i}}$ is the Fourier transformed of the real-space electric field $\vec{E}(\vec{r})$, and the subscript $c$ denotes connected part of the correlation function. The details of the mapping between dimer occupations and electric fields are explained in Section~\ref{sec:suppmat:theory}. Expanding the expression yields
\begin{equation}
    S_{\alpha\alpha}(\mathbf{k}) = \sum_{i,j} \langle \vec{E}_\alpha(\mathbf{r}_i) \vec{E}_\alpha(\mathbf{r}_j)\rangle_c  e^{-i\mathbf{k}\cdot(\mathbf{r}_i - \mathbf{r}_j)},
\end{equation}
which shows that $S_{\alpha\alpha}$ can be directly computed as the 2D Fourier transform of the connected electric field-electric field correlation function. The experimentally obtained electric field correlations and the corresponding structure factors are shown in Fig.~\ref{fig:fig_sm_pinchpoints}. We adopt the same postselection criterion as for the dimer-dimer correlations, considering only links attached to Gauss's-law-satisfying vertices, in addition to postselection on dimer imbalance.

\subsection{Effect of ramp duration on the preparation of QSLs}

\begin{figure*}[h!t!]
\includegraphics{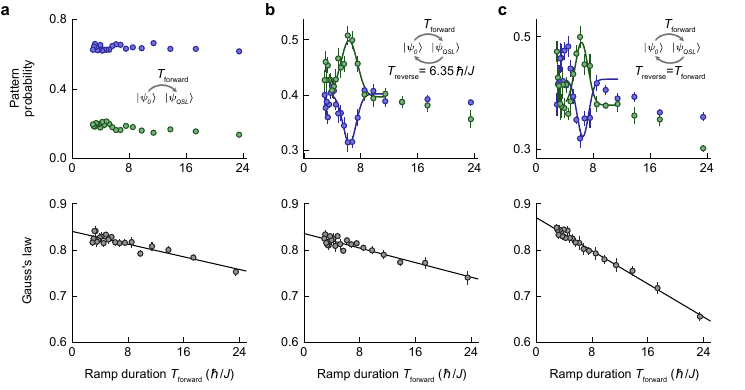}
     \caption{\textbf{Effect of ramp duration on pattern probabilities and Gauss's law validity.}
     \textbf{a,} Detection probabilities of the two Gauss's-law-allowed vertex patterns (single monomer, green; single dimer, blue) measured directly after the forward ramp, as a function of $T_{\mathrm{forward}}$. For $T_{\mathrm{forward}} \gtrsim 2.5\,\hbar/J$ the pattern probabilities reach an approximate plateau, while the fraction of Gauss's-law-allowed vertices (lower panel) decreases with increasing ramp duration.
     \textbf{b,} Post round-trip detection probabilities of the same two patterns as a function of $T_\mathrm{forward}$ with $T_\mathrm{reverse} = 6.35(3)\hbar/J$ fixed. The solid line is a Gaussian fit as a guide to the eye.
     \textbf{c,} Post round-trip detection probabilities of the same two patterns for a symmetric round-trip ($T_\mathrm{forward} = T_\mathrm{reverse}$). The solid line is a Gaussian fit as a guide to the eye. Error bars in all panels denote the s.e.m.
     }
     \label{fig:fig_sm_extended_rampdur_scan}
\end{figure*}

Fig.~\ref{fig:fig_sm_extended_rampdur_scan} shows an extended analysis of the effect of ramp duration $T_\mathrm{forward}$ on the pattern probabilities measured directly after the forward ramp, in a symmetric round-trip experiment ($T_\mathrm{forward} = T_\mathrm{reverse}$), and in a round-trip experiment with the reverse ramp duration fixed ($T_\mathrm{reverse} = 6.35(3) \hbar/J$).

\subsection{Columnar states}
\label{sec:sm_columnar}
In this section, we explore columnar states, which are prepared by applying an additional chemical potential $\mu = \Delta_y - \Delta_x$ during the forward ramp from a monomer initial state, breaking the symmetry between the two lattice axes. Specifically, for $\mu < 0$, the staggered potential $\Delta_y$ is reduced to $\Delta_y + \mu$ while $\Delta_x$ is held fixed, and for $\mu > 0$, the staggered potential $\Delta_x$ is reduced to $\Delta_x - \mu$ while $\Delta_y$ is held fixed.

\begin{figure}[h!]
\includegraphics{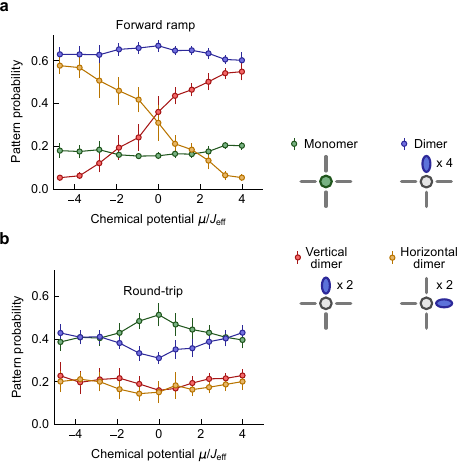}
     \caption{\textbf{Preparation of columnar dimer states.}
        \textbf{a,} Pattern probabilities after the forward ramp as a function of the additional chemical potential $\mu$ applied during the forward ramp.
        \textbf{b,} Same as in (a) for the round-trip experiment.  Each data point was obtained from 14 experimental realisations. The error bars denote the standard deviation obtained using bootstrapping. If not visible, they are smaller than the markers.
     }
     \label{fig:fig_sm_columnar_preparation}
\end{figure}

The probabilities of detecting dimers on horizontal and vertical links measured directly after the forward ramp as a function of $\mu$ are shown in Fig.~\ref{fig:fig_sm_columnar_preparation}a. As expected, a non-zero $\mu$ leads to an imbalance in the occupation of horizontal and vertical links.

The pattern probabilities after the round-trip ($T_\mathrm{forward} = T_\mathrm{reverse}$) are shown in Fig.~\ref{fig:fig_sm_columnar_preparation}b. We find that the probability of detecting vertices occupied by monomers is highest for the balanced state after the forward ramp and decreases with increasing dimer imbalance. This, together with the findings presented in the main text, supports the use of monomer filling after the round-trip as a proxy for the many-body coherence of the QSL.

\begin{figure*}[h!]
\includegraphics{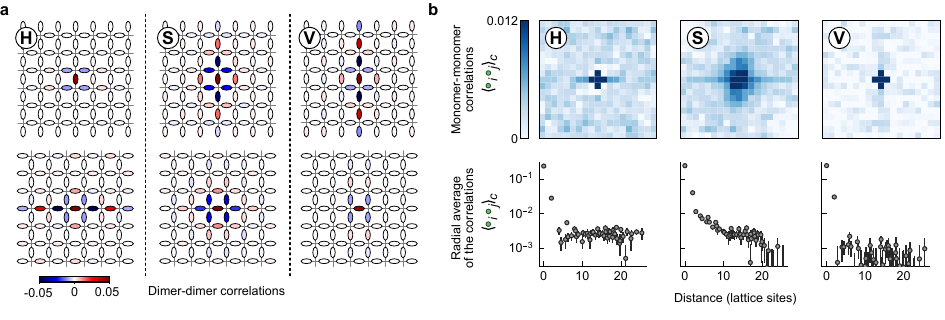}
     \caption{\textbf{Dimer-dimer and monomer-monomer correlations for the columnar states.}
     \textbf{a,} Dimer-dimer correlations after the forward ramp measured for horizontal (H) and vertical (V) columnar states, as well as for the balanced state with symmetrically occupied horizontal and vertical links (S). 
     \textbf{b,} Monomer-monomer correlations after the round-trip experiment for the same states as in (a). The upper three plots show the 2D map of the correlations, the lower three plot the same data radially averaged on a logarithmic scale. Results are obtained from approximately 150 experimental snapshots in each case. The error bars denote the standard deviation obtained using bootstrapping. If not visible, they are smaller than the markers.
     }
     \label{fig:fig_sm_columnar_correlations}
\end{figure*}

We next select two values of the chemical potential, $\mu/J_{\mathrm{eff}} \approx -4.7$ and $\mu /J_{\mathrm{eff}} \approx 4$, together with the balanced case $\mu/J_{\mathrm{eff}} \approx 0$, and show the corresponding dimer-dimer correlations after the forward ramp alongside the monomer-monomer correlations after the round-trip in Fig.~\ref{fig:fig_sm_columnar_correlations}. For the horizontal columnar state, we observe markedly stronger dimer-dimer correlations along the $x$-axis compared to the balanced dimer state, with correlations along the $y$-axis being nearly absent. Conversely, the vertical columnar state exhibits strong correlations along the $y$-axis. Notably, the monomer-monomer correlations are strongly suppressed for both columnar states, despite the relatively modest reduction in monomer filling after the round-trip seen in Fig.~\ref{fig:fig_sm_columnar_preparation}b.

\begin{figure}[h!]
\includegraphics{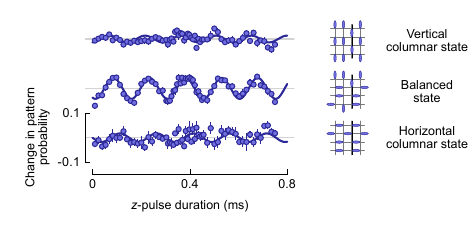}
     \caption{\textbf{$z-$phase imprinting on columnar states.} Oscillations of the dimer-pattern probability as a function of the imprinted phase before the reverse ramp for vertical/horizontal columnar states and the balanced case. Error bars denote one standard deviation from bootstrapping.
     }
     \label{fig:fig_sm_columnar_oscillations}
\end{figure}

Having prepared the columnar states, we proceed to imprint a $z$-phase between the horizontal and vertical dimers, following the same protocol as in Fig.~\ref{fig:coherences} of the main text (Fig.~\ref{fig:fig_sm_columnar_oscillations}). We observe that the amplitude of the dimer pattern oscillations is strongly suppressed for the columnar states compared to the balanced state.

\subsection{Length scale of the QSL regions}

In this section, we describe the extraction of the QSL region length scale using a subsystem analysis, and compare the region size extracted from round-trip snapshots with that extracted from snapshots after the forward ramp.\\

\textbf{Length scale of monomer patches.}

To extract the length scale of the monomer patches after the round-trip experiment, we evaluate the subsystem return probability from the experimental snapshots. As discussed in the main text, the normalised subsystem return probability $-\ln\mathcal{L}_A/A$ saturates to a constant once the subsystem area $A$ exceeds the crossover area $A_c$ that contains all relevant many-body correlations. Intuitively, this means that snapshots in the thermodynamic limit can be assembled from independent patches of area $A_c$ without loss of information.

To extract $A_c$ from $-\ln\mathcal{L}_A/A$, we fit the empirical function
\begin{equation}
    f(A;\,c_0, c_1,\eta)=c_0+c_1\exp(-\eta A),
\end{equation}
and define $A_c$ as the value for which $f(A_c)=1.05\,c_0$.

\begin{figure}[h!]
\includegraphics{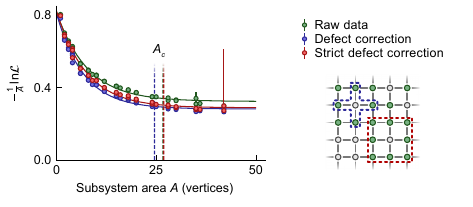}
     \caption{\textbf{Subsystem return probability after the round-trip.} Subsystem return probability $\mathcal{L}_A$ evaluated from round-trip snapshots as a function of subsystem area $A$, for the raw data and after applying two defect-correction criteria. Each data point corresponds to a rectangular subsystem of fixed shape, computed from 877 experimental snapshots. Quasi-1D rectangles whose sides differ by more than 3 vertices are excluded. Error bars are estimated using standard error of proportion; where not visible, they are smaller than the markers.}
     \label{fig:fig_sm_srp_corrections}
\end{figure}

In Fig.~\ref{fig:fig_sm_srp_corrections}, we compare results from the raw snapshots with those obtained after applying two defect-correction criteria, which help mitigate the effect of atom loss during imaging: a single monomer lost from a large patch would otherwise invalidate the entire patch. The correction used in the main text treats empty vertices surrounded by four occupied nearest-neighbour vertices as occupied by monomers. A stricter criterion additionally requires the four next-nearest (diagonal) neighbouring vertices to be occupied (see right inset of Fig.~\ref{fig:fig_sm_srp_corrections}). In all three cases we obtain consistent estimates of $A_c$.\\

\textbf{Length scale of dimer patches.}

\begin{figure}[h!]
\includegraphics{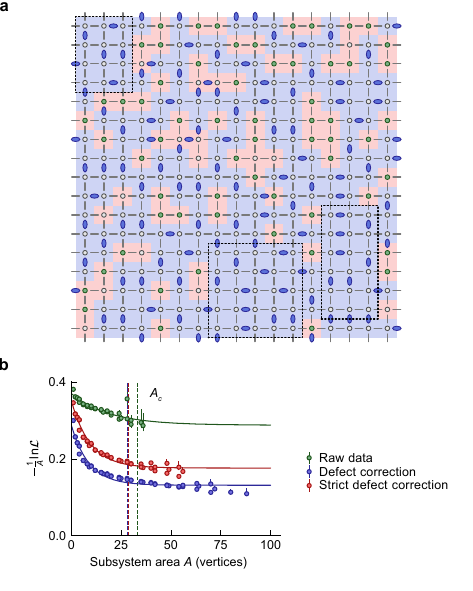}
     \caption{\textbf{Size of regions with near-perfect dimer coverings.}
     \textbf{a,} Vertices with exactly one dimer attached and no monomers ($\langle \hat{P}_V^D\rangle = 1$) are highlighted in blue.
     \textbf{b,} Normalised probability $-\frac{1}{A}\ln\mathcal{L}_A$ as a function of subsystem area $A$, for the raw data and after applying the two defect corrections. Dashed lines mark the extracted crossover area $A_c$. Error bars are estimated using standard error of proportion; where not visible, they are smaller than the markers.
     }
     \label{fig:fig_sm_dimer_patch_analysis}
\end{figure}

Snapshots taken directly after the forward ramp can be analysed in a similar spirit to the subsystem return probability. Instead of projecting onto the initial monomer state, we define
\begin{equation}
    \mathcal{L}_A = \left\langle\prod_{V\in A} \hat{P}_{V}^D\right\rangle,
\end{equation}
where $\hat{P}_V^D$ projects onto the dimer sector at vertex $V$: it equals one if $V$ has exactly one dimer attached and no monomers, and zero otherwise. Fig.~\ref{fig:fig_sm_dimer_patch_analysis}a shows an illustrative snapshot with every vertex labelled by $\langle \hat{P}_V^D\rangle$.

In Fig.~\ref{fig:fig_sm_dimer_patch_analysis}b we plot $-\ln\mathcal{L}_A/A$ and compare results from the raw snapshots with those obtained after applying the same two corrections as above. Here, the corrections check whether the neighbouring vertices belong to the dimer sector rather than whether they are occupied by monomers. The crossover area $A_c$ is extracted using the same procedure as for the monomer patches, and in all three cases, we obtain consistent estimates of $A_c$.\\

\textbf{Extraction of near-perfect dimer and monomer regions in the main-text snapshots.}

Here we describe the procedure used to mark up near-perfect dimer regions in Figs.~\ref{fig:sketch} and~\ref{fig:rk_experiment} and monomer patches in Fig.~\ref{fig:coherences}.

\begin{figure}[t!]
     \includegraphics{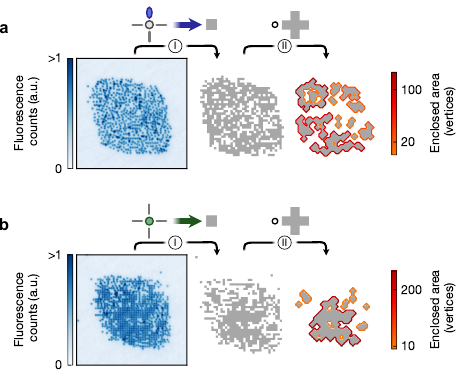}
     \caption{\textbf{Dimer and monomer patches.}
     \textbf{a,} Extraction of near-perfect dimer regions after the forward ramp. Single-site occupations are reconstructed from the fluorescence images (left). Vertices with exactly one dimer attached and no monomers are identified (centre). Vertices surrounded by four nearest-neighbour vertices of the same type are assigned to the bulk, and contours are drawn around the binary-dilated patches (right).
     \textbf{b,} Analogous analysis for monomer patches after the round-trip.}
     \label{fig:SI_sketch_patches}
\end{figure}

Each vertex $V$ is labelled by the value of the projector $\hat P_V^D$ onto the dimer sector, or $\hat P_V$ onto the monomer initial state, as defined above. Vertices with $\hat P_V^{(D)} = 1$ are assigned to be in bulk if all four nearest-neighbour vertices also satisfy $\hat P_V^{(D)} = 1$ (\textit{binary erosion}). The bulk region is then extended by one vertex at the boundary (\textit{binary dilation}), and a contour is drawn around it. For clarity, only the largest domains are shown in Figs.~\ref{fig:sketch} and~\ref{fig:coherences}. The procedure is illustrated in Fig.~\ref{fig:SI_sketch_patches}.

\vspace{0.5em}

\begin{center}
\textbf{SUPPLEMENTARY REFERENCES}
\end{center}
\vspace{0.5em}

\putbib[manuscript]
\end{bibunit}
\end{document}